\documentclass{article}[12pt]
\usepackage{jheppub}
\usepackage[utf8]{inputenc}
\usepackage{amsmath}
\usepackage{amssymb}
\usepackage{amsfonts}
\usepackage{amsthm}
\usepackage{xcolor}
\usepackage{rotating}
\usepackage{array}
\usepackage{appendix}

\title{Superconformal Blocks: General Theory}

\preprint{DESY 19-057, NORDITA 2019-032}

\author{Ilija Buri\' c$^1$,}
\author{Volker Schomerus$^1$}
\author{and Evgeny Sobko$^2,^3$}
\affiliation{$^1$DESY, Notkestra\ss e 85, D-22607 Hamburg, Germany}
\affiliation{$^2$School of Physics and Astronomy, University of Southampton,\\
	Highfield, Southampton, SO17 1BJ, United Kingdom}
\affiliation{$^3$ Nordita, Roslagstullsbacken 23, SE-106 91 Stockholm, Sweden}

\abstract{In this work we launch a systematic theory of superconformal blocks for four-point
	functions of arbitrary supermultiplets. Our results apply to a large class of
	superconformal field theories including 4-dimensional models with any number
	$\mathcal{N}$ of supersymmetries. The central new ingredient is a universal construction
	of the relevant Casimir differential equations. In order to find these equations, we
	model superconformal blocks as functions on the supergroup and pick a distinguished
	set of coordinates. The latter are chosen so that the superconformal Casimir operator
	can be written as a perturbation of the Casimir operator for spinning bosonic blocks
	by a fermionic (nilpotent) term. Solutions to the associated eigenvalue problem can
	be obtained through a quantum mechanical perturbation theory that truncates at
	some finite order so that all results are exact. We illustrate the general theory
	at the example of $d=1$ dimensional theories with $\mathcal{N}=2$ supersymmetry
	for which we recover known superblocks. The paper concludes with an outlook to
	4-dimensional blocks with $\mathcal{N}=1$ supersymmetry.}

\begin{document}

\maketitle

\vspace{1cm}

\addtolength{\baselineskip}{1mm}
\newpage
\section{Introduction}

Conformal blocks are an indispensable tool in conformal field theory that
allows to decompose correlation functions of local and non-local operators into
kinematically determined building blocks with dynamically determined coefficients.
Such decompositions were introduced in the early days of conformal field theory,
but the mathematical nature of the blocks remained elusive for a long time. In
fact, before the work of Dolan and Osborn \cite{Dolan:2000ut,Dolan:2003hv}, the 
shadow formalism of Ferrara et al. \cite{Ferrara:1972uq}, only provided certain 
integral expressions. Dolan and Osborn characterized the blocks through second 
order Casimir differential equations and thereby managed to uncover many of 
their properties, including  explicit formulas for scalar four-point blocks 
in even dimensions. Boosted by the success of the numerical bootstrap 
\cite{Rattazzi:2008pe,ElShowk:2012ht,El-Showk:2014dwa,Simmons-Duffin:2015qma,Kos:2016ysd}, 
several groups contributed to the theory of conformal blocks for scalar and spinning 
fields \cite{Dolan:2011dv,Costa:2011dw,SimmonsDuffin:2012uy,Hogervorst:2013sma,Penedones:2015aga,Costa:2016xah}. In $d=4$ dimensions, for example, conformal blocks 
for correlation functions of four fields with arbitrary spin are known
\cite{Echeverri:2016dun,Cuomo:2017wme}.

In \cite{Isachenkov:2016gim} the authors noticed that the Casimir equations of 
Dolan and Osborn could be considered as stationary Schroedinger equations for 
integrable two-particle Hamiltonians of Calogero-Sutherland type. The latter have 
been studied extensively in the mathematical literature, starting with the work of Heckman
and Opdam \cite{Heckman-Opdam}. The connection uncovered in \cite{Isachenkov:2016gim} 
implies that conformal blocks are certain multivariable hypergeometric functions with
properties that follow from integrability of the quantum mechanical system and
are very similar to those of ordinary hypergeometrics in one variable. Later, it
was shown that the relation between conformal blocks and Calogero-Sutherland
models is universal in that it applies to spinning and defect blocks etc, see  
\cite{Schomerus:2016epl,Schomerus:2017eny,Isachenkov:2018pef}.
\medskip

Except for the conformal window of QCD, whose precise extent is still debated, all
the $d=4$ dimensional conformal field theories we know about are supersymmetric.
The landscape of these superconformal models in higher dimensions is extremely
rich and there exist a variety of complementary techniques to study these theories,
including the AdS/CFT correspondence, localization, non-perturbative dualities and
chiral algebras, to name just a few. Also the bootstrap programme, both in its numerical
and analytical incarnation, has been applied to higher dimensional superconformal
field theories, see \cite{Beem:2013qxa,Beem:2014zpa,Beem:2016wfs}. Significant 
progress has been made by studying four-point blocks of half-BPS operators 
\cite{Dolan:2001tt,Dolan:2004mu,Nirschl:2004pa,Fortin:2011nq,Fitzpatrick:2014oza,
	Bissi:2015qoa,Lemos:2015awa,Liendo:2016ymz,Lemos:2016xke,Chang:2017xmr,Liendo:2018ukf}, 
or superprimary components of correlators for operators that are not half-BPS 
\cite{Khandker:2014mpa,Berkooz:2014yda,Li:2016chh,Li:2017ddj}. For correlation 
functions of BPS operators, the conformal blocks are similar to those of the 
ordinary bosonic conformal symmetry and hence they are well known. On the other 
hand, very little is actually known about blocks for external fields that belong 
to long multiplets. Also the methods developed in \cite{Doobary:2015gia,
	Aprile:2017bgs} have never been applied to such cases. 

The first time the bootstrap programme has been implemented for long multiplets was
in \cite{Cornagliotto:2017dup} for a study of 2-dimensional $\mathcal{N} = (2,0)$
theories (afterwards also in \cite{Kos:2018glc}). In this case blocks for external 
long multiplets could be constructed\footnote{The same methods were recently applied 
	in \cite{Ramirez:2018lpd} to construct blocks for non-BPS operators in 4-dimensional 
	$\mathcal{N}=1$ superconformal field theories.} and it was shown that the numerical 
bootstrap with long multiplets is significantly more constraining than for external 
BPS operators. 
This may not come as a complete surprise: In the example of the 3-dimensional Ising model, 
very strong constraints on the precise location of the model were established by applying 
the numerical bootstrap programme to mixed correlators, i.e.\ to correlators involving two 
different scalar fields. In a supersymmetric theory, long multiplets contain a large 
number of fields so that their correlation functions are highly mixed. Up to now the 
only way to set up a numerical bootstrap for such correlators was through a decomposition 
with respect to blocks of the bosonic conformal group, but since this misses all the 
constraints from supersymmetry, the number of parameters is usually too large to obtain 
good numerical constraints. Superconformal blocks for external long multiplets include 
all the constraints from supersymmetry and hence a long multiplet bootstrap is expected to
incorporate the constraining power of mixed correlators while still allowing for high 
precision numerics. In particular, interesting constraints on superconformal theories 
are expected to arise from the study of four point functions of non-BPS multiplets 
that contain conserved currents or the stress-tensor.

While there are many good reasons to study blocks for long multiplets,
existing results are very sporadic. Our goal is to address this important
issue, in particular for 4-dimensional superconformal field theories. In
this work we shall develop a general algorithm that allows to construct
superblocks for long and short external states systematically, starting
from expressions for spinning bosonic blocks. Our constructions apply
to a large class of superconformal algebras that include all 4-dimensional
ones. Such applications in particular to 4-dimensional theories with
$\mathcal{N}=1$ supersymmetry will be worked out in a forthcoming paper.
\medskip

Let us now summarize the main results of this work and outline the content of the
following sections.

The superprimary fields of a d-dimensional superconformal field theory are associated 
with representations of the group $K = K_\text{bos} \times U$ where $K_\text{bos} = 
SO(1,1) \times SO(d)$ is generated by dilations and rotations and $U$ is a (bosonic) 
group of internal symmetries that includes all $R$-symmetries. We shall denote a finite 
dimensional representation of $K$ on the vector space $V$ by $\pi$.

In the next section we shall explain how to realize superconformal blocks, or rather the associated partial waves (which are sums of the blocks and their shadows) for four-point functions of local operators as functions on a double coset of the bosonic conformal group $G_\text{bos} = SO(1,d+1)$ by the subgroup $K_\text{bos}$ defined above. To each superprimary of weight $\Delta_i$ and spin $\lambda_i$ we associate a finite dimensional representation $\pi_i$ of $K$ on the vector space $V_i$. Given four such fields we shall show that conformal four-point partial waves span the space of functions
\begin{equation}\label{eq:Gamma}
\Gamma = C^\infty\left(K_\text{bos} \backslash G_\text{bos} / K_\text{bos}, W\right)
\end{equation}
in two variables parametrizing the double coset, which take values the finite dimensional vector space
\begin{equation} \label{eq:W}
W = \left( \Lambda \mathfrak{g}_{(1)} \otimes V_1 \otimes \tilde V_2 \otimes V_3
\otimes \tilde V_4\right)^B\ .
\end{equation}
Here $\Lambda \mathfrak{g}_{(1)}$ is the algebra generated by fermionic coordinates on the conformal supergroup $G$ or, in more mathematical terms, the exterior algebra of the vector space of odd elements in the superconformal algebra. This exterior algebra carries a representation of the even subalgebra and hence of the group $K$. The same is true for the spaces $V_i$ and $\tilde V_i$. The group $K$ contains a subgroup $B = SO(d-2) \times U$ and when we build the target space $W$ of our functions we are instructed look for the subspace of $B$-invariant vectors within the tensor product of $K$-representations
that appears inside the brackets on the right hand side of eq.\ \eqref{eq:W}.

The dimension $M = \text{dim} W$  of the vector space $W$ is often referred to as the number of four-point tensor structures in the conformal field theory literature. In the special case that all the external fields are scalars, i.e.\ the vector spaces $V_i$ are 1-dimensional, the number $M-1$ counts the number of so-called nilpotent conformal invariants. Our derivation of the formula \eqref{eq:Gamma} in section 2 requires to extend a theorem for the tensor product of principal series representations of the conformal algebra from \cite{Dobrev:1977qv} to superconformal algebra $\mathfrak{g}$. The analysis in section 2 is similar to the discussion in \cite{Schomerus:2016epl,Schomerus:2017eny} except that we will adopt an algebraic approach and characterize functions on the group through their infinite sets of Taylor coefficients at the group unit. That has the advantage that we
can treat bosonic and fermionic variables on the same footing. Only at the very
end we shall pass from the sets of Taylor coefficients back to functions, at least
for the bosonic variables. The fermionic coordinates remain Taylor expanded, i.e.\
we think of a function on the supergroup and its cosets in terms of the component
functions on the bosonic group that multiply various products of fermionic variables.
\medskip

Even though this different treatment of bosonic and fermionic coordinates breaks
the manifest supersymmetry, our construction of the space \eqref{eq:Gamma} in
section 2 will show that the Laplace operator on the superconformal group $G$,
as well as the higher order operators that are associated with higher Casimir
elements, descend to some well-defined differential operators on $\Gamma$. In
general, these Casimir operators on $\Gamma$ are not that easy to work out and
they may have a complicated form. For superconformal algebras of type I, however,
not only are we able to construct them explicitly but they also turn out as
simple as one could have hoped for. This simplicity results from a very special
choice of coordinates and it makes it possible to construct solutions of the
corresponding eigenvalue problem systematically through a perturbative
expansion that truncates at some finite order and hence gives exact
expressions for superconformal partial waves.

The precise definition of type I superalgebras will be given in section 3.1.
One consequence is the existence of a $U(1)$ $R$-charge, which of course
can be part of a larger $R$-symmetry group. In particular, all superconformal
algebras in $d=4$ dimensions are of type I. The $U(1)$ $R$-symmetry
gives rise to a natural split of the fermionic supercharges $Q$ into two set
$Q^\pm_\alpha$ of opposite $R$-charge and similarly for the special
superconformal transformations $S$. We use this split when we introduce
coordinates on the supergroup $G$ as
\begin{equation} \label{eq:variables}
G =  e^{\bar\sigma^\beta_q Q^-_\beta + \bar \sigma_s^\beta S^-_\beta} \ G_{(0)} \
e^{\sigma^\beta_q Q^+_\beta + \sigma_s^\beta S^+_\beta}\ .
\end{equation}
Here, $G_{(0)}$ is the bosonic subgroup of $G$ and $\sigma_{q,s}^\alpha,
\bar \sigma_{q,s}^\alpha$ are Grassmann valued coordinates. Using a result from
\cite{Quella:2007hr} we are then able to show that the Casimir equation for
superconformal groups of type I takes the form of an eigenvalue equation for
a Calogero-Sutherland Hamiltonian of the form
\begin{equation} \label{eq:Casimir}
H = H_0 + A = H_0 - 2D^{ab}(z,\bar z)\bar \partial_a \partial_b \ .
\end{equation}
Here $H_0$ is the Calogero-Sutherland Hamiltonian that gives rise to the
Casimir equations of a specific set of spinning conformal blocks for the
bosonic group $G_{(0)}$. These are Calogero-Sutherland Hamiltonians with a
matrix valued potential whose form was worked out in \cite{Schomerus:2016epl,Schomerus:2017eny}. The bosonic Hamiltonian $H_0$ is perturbed by a second term
$A$ that may be considered as a nilpotent matrix valued potential. The
operators $\bar \partial_a$ and $\partial_b$ denote derivatives with respect to the
fermionic coordinates. We can think of them as annihilation operators on
the exterior algebra $\Lambda \mathfrak{g}_{(1)}$ and hence may represent
them as matrices on the tensor product of spaces that appears inside the
brackets on the right hand side of eq.\ \eqref{eq:W}. The combination
of annihilation operators that appears in the second term of $H$ is
$B$-invariant and hence acts on $\Gamma$ as a matrix valued potential
term. The matrix elements  are functions of the bosonic cross ratios
$z,\bar z$ that are quite easy to work out explicitly.

There are two remarkable aspects about this result for the Casimir
operator on $\Gamma$. The first concerns the fact that we were able
to split off the bosonic Casimir operator for spinning blocks. The
associated Casimir equations are actually well studied at least for
dimension $d\leq 4$. In particular, spinning bosonic blocks for $d=4$
are known from \cite{Echeverri:2016dun}. The second remarkable fact concerns the form
of $A$: Naively one might have expected that
the coefficient functions $D^{ab}$ that multiply the annihilation
(derivative) operators depend on nilpotent invariants as well. These
would have been interpreted as creation operators on the exterior
algebra. But this is not the case so
that the second term $A$ in our Hamiltonian is nilpotent, i.e.\
$A^N = 0$ for some integer $N$. Since the eigenvalue problem for
$H_0$ is solved, we want to consider $A$ as a perturbation.
Nilpotency of $A$ guarantees that the quantum mechanical perturbation
theory truncates at order $N-1$ so that we can obtain exact results
by summing just a few orders of the perturbative expansion. It turns
out that, at any order of the expansion, the perturbation may be
evaluated explicitly with some input from the representation theory
of $SO(d+2)$. In this sense our results provide a complete solution
of the Casimir equations for type I superconformal symmetry and in
particular for 4-dimensional conformal field theories with any
number $\mathcal{N}$ of supersymmetries.

In section 4 we will illustrate this programme through the example
of $\mathcal{N}=2$ superconformal symmetry in $d=1$. The corresponding
conformal blocks were originally constructed in \cite{Cornagliotto:2017dup}
and here we shall show how to recover these results by summing two orders
of the quantum mechanical perturbation theory we described above. While
most of our analysis is motivated by the desire to construct conformal
blocks for four-point functions of long (non-BPS) multiplets, we shall
also discuss how to implement shortening conditions, both in general and
in our $d=1$ dimensional example. The paper concludes with a short outlook
on the construction of blocks in 4-dimensional $\mathcal{N}=1$ theories.
A detailed analysis of such blocks is in preparation, at least for
four-point functions of two long and two BPS operators.

\section{Partial Waves from Harmonic Analysis}

\def\Lie{\mathit{Lie}}

Following the strategy laid out in \cite{Schomerus:2016epl}, our first task is to realize superconformal partial waves\footnote{We recall that a conformal partial wave is a sum of a block and its shadow. Our discussion in the next sections focuses on partial waves. The decomposition into blocks can be performed at the end.} as functions on the superconformal group. The observation which allows us to generalize many of the results from \cite{Schomerus:2016epl} to supersymmetric setup is that one can pass from the notion of an induced representations of a Lie group to that of a coinduced representation of its Lie algebra. Once this is done, adding supersymmetry largely amounts to inserting appropriate minus signs into the formulas. We shall initially consider correlators of generic (long) multiplets and then discuss the effect of shortening conditions in the final subsection.

\subsection{Group theoretic background}

Before we can start our discussion of conformal partial waves, we need to introduce a bit
of notation concerning the superconformal Lie algebra and the associated group. Let $G$ be
some superconformal group, and $\mathfrak{g} = \Lie(G)$ the corresponding superconformal
algebra. As usual, $\mathfrak{g}$ splits into an even and an odd part. The even part
$\mathfrak{g}_{(0)}$ is assumed to contain the conformal Lie algebra $\mathfrak{so}(1,d+1)$
as a direct summand. The other summands in $\mathfrak{g}_{(0)}$ describe internal symmetries.
We shall denote these summands by $\mathfrak{u}$. In addition to the standard $\mathbb{Z}_2$
grading that distinguishes between even and odd elements, the conformal Lie superalgebra
$\mathfrak{g}$ admits a finer grading which comes from decomposition into eigenspaces of
$\text{ad}_D$, where $D$ is the generator of dilations,
\begin{equation}
\mathfrak{g} = \mathfrak{g}_{-1}\oplus\mathfrak{g}_{-1/2} \oplus \mathfrak{g}_0
\oplus\mathfrak{g}_{1/2}\oplus\mathfrak{g}_{1} =
\tilde{\mathfrak{n}} \oplus \mathfrak{s} \oplus \mathfrak{k} \oplus \mathfrak{q}
\oplus \mathfrak{n} . \nonumber
\end{equation}
The even part of $\mathfrak{g}$ is composed of $\mathfrak{g}_{\pm 1}$ and $\mathfrak{k}$ where
$\mathfrak{g}_{-1}= \tilde{\mathfrak{n}}$ contains the generators $K_\mu$ of special conformal
transformations while $\mathfrak{g}_{1} = \mathfrak{n}$ is generated by translations $P_\mu$.
Dilations, rotations and internal symmetries make up
$$ \mathfrak{k} = \mathfrak{so}(1,1) \oplus \mathfrak{so}(d) \oplus \mathfrak{u} \ . $$
Generators of $\mathfrak{g}_{\pm1/2}$, finally, are supertranslations $Q_\alpha$ and super
special conformal transformations $S_\alpha$. We shall also denote these summands as $\mathfrak{s}
= \mathfrak{g}_{-1/2}$ and $\mathfrak{q} = \mathfrak{g}_{1/2}$.

The even and the odd subspace $\mathfrak{g}_{(0)}$ and $\mathfrak{g}_{(1)}$ both carry a
representation of the even subalgebra $\mathfrak{g}_{(0)}$ (and hence, also its subalgebra
$\mathfrak{k}$) which is defined through the adjoint action. We will denote by $\Lambda
\mathfrak{g}_{(1)}$ the exterior algebra of the odd subspace. It inherits a $\mathfrak{g}_{(0)}$-module
structure from that on $\mathfrak{g}_{(1)}$ in a natural way. In particular the exterior algebra
carries a representation of the Lie algebra $\mathfrak{k}$. From the Poincare-Birkhoff-Witt-theorem for Lie superalgebras, the universal enveloping algebra of $\mathfrak{g}$ is expressible as the vector space tensor product
\begin{equation} \label{eq:UU0L}
U(\mathfrak{g}) \cong \Lambda\mathfrak{g}_{(1)}  \otimes U(\mathfrak{g}_{(0)})\ .
\end{equation}
One may think of the  universal enveloping algebra as the algebra of all right or left invariant
differential operators on the supergroup. If we remove all the differential operators
with respect to bosonic coordinates from the algebra of all differential operators, we remain
with differential operators on the fermionic coordinates, i.e.\ with the exterior algebra.

Both tensor factors on the right hand side of \eqref{eq:UU0L} come equipped with
an action of the even subalgebra $\mathfrak{g}_{(0)}$ that is inherited from the adjoint
action on $U(\mathfrak{g})$. The isomorphism \eqref{eq:UU0L} respects this action, i.e.\
it is an isomorphism of $\mathfrak{g}_{(0)}$ modules under the adjoint action. If one
replaces the adjoint action of $\mathfrak{g}_{(0)}$ on $U(\mathfrak{g})$ and
$U(\mathfrak{g}_{(0)})$ with the left or right regular action the isomorphism above
remains valid. It will be of some importance, however, that these isomorphisms are not canonical.

As a final element of notation, let us also agree to use $\mathfrak{p}$ for the subalgebra
that is generated by vectors of degree less than or equal to zero,
\begin{equation}
\mathfrak{p} = \mathfrak{g}_{-1}\oplus\mathfrak{g}_{-1/2}\oplus\mathfrak{g}_0 \nonumber
\end{equation}
and call it the {\it parabolic subalgebra}. There is a unique (connected) corresponding
subgroup $P\subset G$ such that $\mathfrak{p} = \Lie(P)$. The superspace can be identified
with the group of translations and supertranslations. It is defined as the homogeneous
space $G/P = \overline{\mathbb{C}^{d|k}}$. For more details about supermanifolds, Lie
supergroups and their homogeneous spaces, we refer to the classical work of Kostant,
\cite{Kostant:1975qe}.

\subsection{Functions on supergroups}

In \cite{Schomerus:2016epl,Schomerus:2017eny} all relevant representations of the conformal
symmetry and ultimately the space of conformal partial waves were realized as functions on
the conformal group. The notion of a function on a supergroup is a bit more subtle and for this reason we will adopt a more algebraic approach. In brief, we shall pass from a function on the group to an infinite set of its Taylor coefficients at the unit element. It is intuitively clear that these should contain the same information as the function itself, but the generalization to supergroups turns out to be straightforward.

For a moment let us assume that $G$ is an ordinary bosonic connected group and consider
the space
\begin{equation} \label{eq:funct}
\Gamma_G^V := C^\infty (G,V)
\end{equation}
of smooth functions on $G$ taking values in some vector space $V$. This space of functions
carries two commuting actions of the Lie algebra $\mathfrak{g}$ by left and right invariant
vector fields. In other words, there exist maps
$$ x \mapsto \mathcal{L}_x\ , \quad x \mapsto \mathcal{R}_x \  $$
that send arbitrary elements $x \in \mathfrak{g}$ to first order differential operators
on the space of functions such that the Lie bracket it realized as the commutator. From
the first order operators we can build up higher order ones by talking products and sums.
In this way we obtain the algebra of left (resp. right) invariant differential operators on
the space of functions. The latter may be thought of as a realization of the universal
enveloping algebra $U(\mathfrak{g})$.

Given a function $f \in C^\infty(G,V)$ and an element $A \in U(\mathfrak{g})$ we can
compute the vector
$$ \varphi_f (A) := \mathcal{L}_A f (e) \in V , $$
i.e.\ we can evaluate the action of the differential operator $\mathcal{L}_A$ on $f$
at the group unit $e \in G$. As we scan over all elements $A\in U(\mathfrak{g})$, i.e.\
all left invariant differential operators, we obtain all the coefficients of the Taylor
expansion at the group unit. In this way, we store the information of $f$ in a linear
map
\begin{equation} \label{eq:Taylor}
\varphi = \varphi_f \in \Theta_G^{V} := \text{Hom}(U(\mathfrak{g}),V) \ \ .
\end{equation}
Just as the space $C^\infty(G,V)$ of functions $f$ on the group carries two commuting
actions of the Lie algebra $\mathfrak{g}$, so does the linear space $\Theta_G^V$  of
Taylor coefficients. We denote these actions by the same letters $\mathcal{L}$ and
$\mathcal{R}$. For elements $x \in \mathfrak{g}$ they
are given by
\begin{equation}
\mathcal{L}_x \varphi(Y) =  \varphi(Yx) \quad , \quad
\mathcal{R}_x \varphi(Y) = -\varphi(xY) \ .
\end{equation}
Our discussion shows that the space \eqref{eq:funct} of functions on $G$ and the
space \eqref{eq:Taylor} possess the same structure.

Up to this point we have assumed that the Lie algebra $\mathfrak{g}$ is bosonic. Let
us now go back to the case in which $\mathfrak{g}$ is a Lie superalgebra. While it is
not too obvious how too make sense of $C^\infty(G,V)$, at least not without making
some choice of the coordinate system on the supergroup, the space
$\Theta_G^V$ from eq.\ \eqref{eq:Taylor} of Taylor coefficients is well defined for
any $\mathbb{Z}/2$-graded vector space $V$. So, when dealing with superalgebras
$\mathfrak{g}$, we simply adopt the latter as our basic model space, thereby avoiding
all the issues of defining $C^\infty(G,V)$.

For supergroups it is actually quite common to go back from elements $\varphi \in
\Theta_G^V$ to a function on the bosonic group $G_{(0)}$ with the help of the isomorphism
\eqref{eq:UU0L} we discussed in the previous subsection,
\begin{equation}
\text{Hom}\left(U(\mathfrak{g}),V\right)  \cong \text{Hom}\left(\Lambda \mathfrak{g}_{(1)}
\otimes U(\mathfrak{g}_{(0)}),V\right) = \text{Hom} \left(U(\mathfrak{g}_{(0)}),\Lambda
\mathfrak{g}_{(1)}^\ast \otimes V\right)\ .
\end{equation}
In the final step we traded the exterior algebra in the first argument for its dual
in the second. The latter is denoted by $\Lambda \mathfrak{g}_{(1)}^\ast$. The linear
space on the right hand side may now be interpreted as a space of Taylor coefficients
of functions
\begin{equation} \label{eq:GammaVG}
f \in \Gamma_G^V := C^\infty(G_{(0)},\Lambda\mathfrak{g}_{(1)}^\ast\otimes V)\ ,
\end{equation}
i.e.\ of smooth functions on the bosonic group that take values in the tensor product
of $V$ with the dual of the exterior algebra. When we pass from $\varphi \in \Theta_G^V$
to $f \in \Gamma_G^V$, we leave the dependence on fermionic variables Taylor expanded.
The different treatment of fermionic and bosonic variables has the advantage that we
end up with ordinary functions, but it breaks supersymmetry. In particular, we noted
above that isomorphisms in \eqref{eq:UU0L} with respect to the left or right regular
actions of the even subalgebra are not canonical, i.e.\ they require additional choices that
are associated with a choice of fermionic coordinates. Whenever such a choice is made, the
actions are implemented by a set of matrix valued first order differential operators with a
non-trivial action on the exterior algebra. We will adopt such a choice later on in our
discussion of type I superconformal groups, but for the moment we will simply work with
the space $\Theta_G^V$ and various subspaces thereof.

\subsection{Fields and principal series representations}

With all this notation in place, let us now consider some superconformal field theory. Local operators $\mathcal{O}$ in such a theory are labeled by finite dimensional irreducible representations $V$ of $\mathfrak{k}$. The commutation relation of a primary field with the with the (super-)conformal generators mean that fields are associated with principal series representations. In \cite{Schomerus:2016epl,Schomerus:2017eny}
these representations of the conformal algebra were realized in terms of functions of the conformal group satisfying the covariance law
\begin{equation} \label{eq:induced}
f(pg) = \pi (p) f(g) \, \text{for} \, p
\in P \ .
\end{equation}
Here we have extended the finite dimensional irreducible representation $V$ of
$\mathfrak{k}$ to a representation of the parabolic subalgebra $\mathfrak{p}$
by acting trivially with generators of $\mathfrak{p} / \mathfrak{k}$. 

As we explained in the previous subsection, this is not the route we want to take
here. Instead, we will realize the principal series representation on an appropriate subspace of the space $\Theta_G^V$ defined in \eqref{eq:Taylor} which is given
by
\begin{equation}\label{eq:coindW}
\Theta^V_{G/P} = \text{Hom}^{L}_{U(\mathfrak{p})} (U(\mathfrak{g}),V) ,
\end{equation}
i.e.\ we realize the principal series representation on the space of complex linear
maps $\varphi: U(\mathfrak{g}) \xrightarrow{} V$ which satisfy
$$ \varphi(x A) =  (-1)^{|x||\varphi|}\pi(x) \varphi(A)\ \quad \text{for} \quad x \in
\mathfrak{p}\ . $$
Here, and in the rest of the paper, $|.|$ denotes the parity of a homogeneous element in a super vector space. The equation above \textit{defines} the parity of a covariant map $\varphi$.  The superscript $L$ we placed on the $\text{Hom}$ reminds us that
we use the left action
of $U(\mathfrak{p})$ to formulate this condition. The index $G/P$ on
$\Theta$ may seem a bit unnatural since the right hand side only involves $\mathfrak{g}$
or its universal enveloping algebra. Nevertheless we will continue to label the space
$\Theta$ by groups and cosets thereof to keep notations as close as possible to those
used in \cite{Schomerus:2016epl,Schomerus:2017eny}, except that now we use the spaces
$\Theta$ of Taylor expansions rather than spaces $\Gamma$ of functions. For later use
we also note that our definition \eqref{eq:coindW} implies that any element
$\varphi$ of $\Theta$ satisfies
\begin{equation}
\varphi(y U(\mathfrak{g})) =0\quad , \ \textit{for}\ y \in \mathfrak{p}/
\mathfrak{k}\ ,   \label{com}
\end{equation}
as a consequence of the fact that special (super)conformal transformations act
trivially on the space $V$. The action of $x\in\mathfrak{g}$ on maps $\varphi
\in \Theta^V_{G/P}$ is given by
\begin{equation}
(x\varphi)(A) = (-1)^{|x|(|\varphi|+|A|)}\varphi(A x),\ A\in U(\mathfrak{g}). \nonumber
\end{equation}
This action defines a representation of the superalgebra $\mathfrak{g}$ on
$\Theta_V$ that we shall denote by $\pi_V$. For a certain set of
representations $\pi$ of the subalgebra $\mathfrak{k}$, the corresponding
representation $\pi_V$ of the superconformal algebra $\mathfrak{g}$ is said to belong to the \textit{algebraic principal series}. The precise conditions on $V$ are not relevant in
our subsequent discussion. In mathematics, the construction \eqref{eq:coindW}
of the representation space is known as \textit{coinduction} or \textit{production}. 
As we have explained above, coinduction of representations of Lie algebras naturally 
corresponds to induction of associated representations of Lie groups. For some more 
details, see the paper of Blattner, \cite{Blattner}.

\subsection{Tensor products of principal series representations}

The superconformal partial waves we consider in this work are relevant for decompositions of
correlation functions of four local primary operators,
\begin{equation}
\Phi = \langle \mathcal{O}_1 (x_1,\theta_1)...\mathcal{O}_4 (x_4,\theta_4) \rangle,
\nonumber
\end{equation}
Here, the bracket $\langle \cdot \rangle$ denotes the vacuum expectation value which is
$\mathfrak{g}$ invariant by definition. Hence, except for a simple prefactor, the four-point
function $\Phi$ is an element of the space of $\mathfrak{g}$-invariants in the four-fold
tensor product of principal series representations
\begin{equation}
\Phi \ \in \   \Theta =\Big(\bigotimes_{i=1}^4 \Theta_{G/P}^{V_i}\Big)^\mathfrak{g}. \label{inv}
\end{equation}
In order to determine the space $\Theta$ of
these invariants one starts by computing the tensor product of two principal series
representations \cite{Schomerus:2016epl,Schomerus:2017eny}. In the case of bosonic
conformal groups, the relevant mathematical theorem was found in \cite{Dobrev:1977qv},
see theorem 9.2. Here we shall formulate and prove this theorem in the superconformal
setting.

The formulation of our new theorem requires a bit of additional algebraic structure, namely
the so-called \textit{Weyl inversion}, an inner automorphism $s$ of $\mathfrak{g}$ that may
be constructed as follows. The restricted Weyl group of the ordinary conformal group $SO(d+1,1)$
consists of two elements $\{1,\sigma\}$, \cite{Dobrev:1977qv}. The non-trivial element $\sigma$
is an inner automorphism of $\mathfrak{so}(d+1,1)$ that is implemented by conjugation with
an element $w \in G$,
\begin{equation}\label{eq:sinv}
\sigma(x) = \textit{Ad}_w(x) = w x w^{-1}.
\end{equation}
It is well known that $\sigma$ flips the sign of the dilation generator, $\sigma(D)=-D$. We can
now lift the equation \eqref{eq:sinv} to a definition of the Weyl inversion - an automorphism $s$ of
the full superconformal algebra. Since it flips the sign of $D$, we infer that
\begin{equation}
s(\mathfrak{g}_{a}) = \mathfrak{g}_{-a},\
\nonumber
\end{equation}
for $a = 0, \pm 1/2, \pm 1$. In particular, $s$ maps $\mathfrak{k}$ onto itself. Given any
representation $\pi$ of $\mathfrak{g}$, we denote by $\pi^s$ the representation obtained by 
composing it with $s$, i.e.\ we define $\pi^s = \pi\circ s$. Clearly, for the coinduced
module $\pi_V$, i.e.\ the $\mathfrak{g}$ representation on the space $\Theta^V_{G/P}$,
the carrier space of $\pi_V^s$ takes the form
\begin{equation} \label{eq:barcoindW}
\bar \Theta^{V'}_{G/P} :=
\text{Hom}^L_{U(s(\mathfrak{p}))}\left(U(\mathfrak{g}),V' \right)\ ,
\end{equation}
where the representation $V'$ of $s(\mathfrak{p})$ is obtained from the representation $V$ of
$\mathfrak{k}$ by composition with the Weyl inversion and a trivial extension to $s(\mathfrak{p})$.
As we recalled above, the inversion flips the sign of $D$. On the other hand, it acts trivially
on the generators $U \in \mathfrak{u}$ of the internal symmetry and it maps a representation $l$
of the rotation group to its dual $l^\ast$. We will denote the $\mathfrak{g}$ representation on
the space $\bar \Theta^{V'}_{G/P}$ by $\bar\pi_{V'}$. The bar is supposed to remind us that we
perform our coinduction from the image $s(\mathfrak{p})$ of the parabolic subalgebra under the
Weyl inversion. Since $s$ is inner, we of course have $\pi_V\cong\bar\pi_{V'}$ or
$$ \Theta^V_{G/P} \cong \bar \Theta^{V'}_{G/P} \ \  . $$
Now we are prepared to state our theorem. Let us take two principal series
representations associated with the two $\mathfrak{k}$ modules $V_1$ and $V_2$,
respectively. Then the tensor product of these two representations is given by
\begin{equation}\label{eq:superMack}
\Theta^{V_1}_{G/P} \otimes \Theta_{G/P}^{V_2} \cong
\Theta^{V_1 \otimes V'_2}_{G/K} := \text{Hom}^L_{U(\mathfrak{k})}
\left(U(\mathfrak{g}), V_1 \otimes V'_2 \right)\ .
\end{equation}
In order to prove this statement we shall construct the following
linear map $F$ between the modules on the left and the right hand
side,
\begin{align*}
F:& \Theta^{V_1}_{G/P} \otimes \bar \Theta^{V'_2}_{G/P}  \xrightarrow{} \text{Hom}^L_{U(\mathfrak{k})}(U(\mathfrak{g}),
V_1\otimes V'_2),\\[2mm]
& \varphi_1 \otimes \varphi_2  \mapsto \psi = (\varphi_1\otimes\varphi_2)
\circ \Delta.
\end{align*}
Here, the two irreducible representations $\Theta$ and $\bar\Theta$ are defined
as in eqs.\ \eqref{eq:coindW} and \eqref{eq:barcoindW} with $V = V_1$ and $V' =
V_2'$. The symbol $\Delta$ denotes the usual coproduct in $U(\mathfrak{g})$.
Using standard properties of the coproduct it is easy to see that $F$ is well
defined and a homomorphism of $\mathfrak{g}$-modules. Moreover, it is invertible.
To show this, let $\psi$ be an arbitrary element of the module on the right hand side. We shall reconstruct from it a function
\begin{equation}
\varphi:U(\mathfrak{g})\otimes U(\mathfrak{g})\xrightarrow{}V_1\otimes V'_2, \nonumber
\end{equation}
which is its preimage in the representation space on the left. Due to covariance properties, such a function is completely specified by the values
\begin{equation}
\varphi(P_1^{n_1}... P_d^{n_d}Q_1^{\varepsilon_1}...Q_k^{\varepsilon_k}\otimes 1),\ \varphi(1\otimes S_1^{\eta_1}... S_k^{\eta_k}K_1^{m_1}...K_d^{m_d})
\end{equation}
Further, by the equation \eqref{com}
\begin{equation}
\varphi(S_1^{\eta_1}... S_k^{\eta_k}K_1^{m_1}...K_d^{m_d}\otimes 1) = 0 = \varphi(1\otimes P_1^{n_1}... P_d^{n_d}Q_1^{\varepsilon_1}...Q_k^{\varepsilon_k}). \nonumber
\end{equation}
Using that $\Delta(X) = X \otimes 1 + 1 \otimes X$ for all elements $X \in \mathfrak{g}$
we conclude that
\begin{equation}
\psi(P_1^{n_1}... P_d^{n_d}Q_1^{\varepsilon_1}...Q_k^{\varepsilon_k}) =
\varphi(P_1^{n_1}... P_d^{n_d}Q_1^{\varepsilon_1}...Q_k^{\varepsilon_k}\otimes 1),
\nonumber
\end{equation}
and similarly for the second type of elements. Hence, we are able to recover $\varphi$ from $\psi$. This shows that $F$ is an isomorphism of $\mathfrak{g}$-modules and it is a natural generalization
of Theorem 9.2 in \cite{Dobrev:1977qv} to the context of superconformal symmetry.

\subsection{The space of conformal partial waves}

Now that we have realised the carrier space for the tensor product as stated in eq.\
\eqref{eq:superMack}, we are able to compute the space $\Theta$ of superconformal partial
waves,
\begin{eqnarray}
\Theta_{K\backslash G/K}^{V_{(12)},V_{(34)}} & := & \Theta =
\left( \bigotimes_{i=1}^4 \Theta_{G/P}^{V_i}\right)^{\mathfrak{g}} \nonumber \\[2mm]
& = & \left( \text{Hom}^L_{U(\mathfrak{k})}(U(\mathfrak{g}),V_{(12)})
\otimes \text{Hom}^R_{U(\mathfrak{k})}(U(\mathfrak{g}),V_{(34)})\right)^{\mathfrak{g}} \nonumber
\\[2mm]
& = & \left( \text{Hom}^{LR}_{U(\mathfrak{k})\otimes U(\mathfrak{k})}
(U(\mathfrak{g}) \otimes U(\mathfrak{g}),V_{(12)} \otimes V_{(34)}) \right)^{\mathfrak{g}}
\nonumber \\[2mm]
& = & \text{Hom}^{LR}_{U(\mathfrak{k}) \otimes U(\mathfrak{k})} \left(U(\mathfrak{g}),
V_{(12)} \otimes V_{(34)}\right)\ . \label{eq:CPWlong}
\end{eqnarray}
In the first step we have inserted our result \eqref{eq:superMack} to evaluate the
tensor product of $\pi_1 \otimes \pi_2 \cong \pi_1 \otimes \bar \pi_2'$. For the
tensor product of $\pi_3 \otimes \pi_4$ we employed a similar statement using the
right action of $\mathfrak{k}$ on $U(\mathfrak{g})$ instead of the left action, i.e.
\begin{equation}
\Theta^{V_3}_{G/P} \otimes \Theta^{V_4}_{G/P} \cong \text{Hom}^R_{U(\mathfrak{k})}
\left(U(\mathfrak{g}),V_3 \otimes V_4'\right)\ .
\end{equation}
In addition we introduced the shorthands $V_{(12)} = V_1\otimes V_2'$ and $V_{(34)}
= V_3\otimes V_4'$. Then we used
$$\left(U(\mathfrak{g}) \otimes U(\mathfrak{g})\right)^{\mathfrak{g}}
\cong U(\mathfrak{g}) \
$$
to pass to the space of invariants. In the final line, the first tensor factor $U(\mathfrak{k})$ acts on $U(\mathfrak{g})$
from the left and it acts trivially on $V_{(34)}$. The second tensor factor acts from
the right and trivially on $V_{(12)}$. Eq.\ \eqref{eq:CPWlong} is the final results of
our short computation and it provides a concise description of the space of conformal
partial waves.
\medskip

In order to appreciate the results a bit better, we now want to go back from our spaces
$\Theta$ of Taylor expansions to functions. As we have explained in the second subsection,
we shall model the space $\Theta$ in eq.\ \eqref{eq:CPWlong} in terms of functions on the
bosonic group. In order to do so, we make use of the isomorphism \eqref{eq:UU0L} to
write
\begin{eqnarray}
\Theta  & = &  \text{Hom}^{LR}_{U(\mathfrak{k}) \otimes U(\mathfrak{k})}
\left(U(\mathfrak{g}),V_{(12)} \otimes V_{(34)}\right) \nonumber \\[2mm]
& = & \text{Hom}^{LR}_{U(\mathfrak{k}) \otimes U(\mathfrak{k})} \left(
\Lambda \mathfrak{g}_{(1)} \otimes U(\mathfrak{g}_{(0)}), V_{(12)}
\otimes V_{(34)}\right) \nonumber \\[2mm]
& = & \text{Hom}^{LR}_{U(\mathfrak{k}) \otimes U(\mathfrak{k})}
\left(U(\mathfrak{g}_{(0)}), [\Lambda \mathfrak{g}_{(1)}^\ast\otimes V_{(12)}] \otimes
V_{(34)}\right)\ . \label{eq:CPWlongsplit}
\end{eqnarray}
As we explained before, the isomorphism \eqref{eq:UU0L}  with respect to left or 
right action of $U(\mathfrak{g}_{(0)})$ is not canonical. Consequently the extension 
of the left and right action of $U(\mathfrak{k}) \otimes U(\mathfrak{k})$ in the last 
step is possible, but it depends on certain choices. Here we worked with one
such choice in which the right factor $U(\mathfrak{k})$ acts trivially on the
exterior algebra and its dual. In the target space we placed square brackets around
the first two tensor factors to remind us that these are acted upon trivially by
the right $U(\mathfrak{k})$. Our answer seems to break the symmetry between left
and right, but we could have equally well moved the exterior algebra into the right
tensor factor or we could have even split is to put part of it into the left and
the rest into the right. All these spaces are certainly isomorphic as vector spaces. 
Finally, let us note that $\Lambda\mathfrak{g}_{(1)}^\ast\cong\Lambda\mathfrak{g}_{(1)}$ 
as $\mathfrak{g}_0$-modules, so we can safely drop the star in various formulas 
that will follow.
\medskip

It is now possible to interpret the space of superconformal partial waves in terms of
functions on the bosonic group. By comparison with the discussion of bosonic partial
waves in \cite{Schomerus:2016epl,Schomerus:2017eny} we conclude that the corresponding
space of functions is given by
\begin{equation}\label{eq:CPWfunctions}
\Gamma_{K\backslash G /K}^{V_{(12)},V_{(34)}} := C^\infty\left(K\backslash G_{(0)}/K,
\left(\Lambda \mathfrak{g}_{(1)}\otimes V_{(12)}\otimes
V_{(34)}\right)^{B}\right) \ .
\end{equation}
The presence of the stabiliser group $B$ accounts for the fact that left and right actions
of $K$ on the bosonic subgroup $G_{(0)}$ are not independent. Once we divide $G_{(0)}$ by
$K$ from the right, the left action of $K$ on $G_{(0)}/K$ is no longer free, with $B
\subset K$ being the stabilizer subgroup for this action. In eq.\ \eqref{eq:CPWfunctions}
the group $B$ is embedded diagonally into the product $K \times K$ and hence we can drop
the distinction between left and right factor. In other words, the choice we have made
in eqs.\ \eqref{eq:CPWlongsplit} becomes irrelevant in the final result. Compared to the
bosonic case, both the subgroup $K$ and the stabilizer subgroup $B$ are multiplied by the
internal symmetry subgroup $U$,
\begin{equation}
K = K_\textit{bos}\times U\quad ,\ B = B_\textit{bos}\times U. \nonumber
\end{equation}
The second fact comes from the observation that internal symmetries commute with conformal
transformations. The dimension $M$ of the space of $B$ invariants,
\begin{equation} \label{eq:M}
M = \text{dim} \left(V_{(12)} \otimes V_{(34)} \otimes \Lambda
\mathfrak{g}_{(1)}\right)^B\
\end{equation}
computes the number of four-point tensor structures for generic long external
multiplets. Of course, this number may be reduced by shortening conditions and
conservations laws when we insert external BPS fields, conserved currents etc.

\subsection{Shortening conditions}
\label{shortening}

Our discussion so far was restricted to long multiplets. We now extend it to the case in which
some or all of the four fields satisfy additional BPS conditions. This means that the superconformal
primary operators $\mathcal{O}_i$ are annihilated by a certain set $I_i$ of generators $\{Q_\alpha\}$
with $\alpha \in I_i$.\footnote{More generally, shortening conditions can also involve products of
	supercharges. It is straightforward to include such cases into our discussion.} We shall denote the
spaces that are spanned by these generators as
\begin{equation}\label{spaces}
\mathfrak{Q}_i = \textit{span}\{ Q_\alpha, \alpha \in I_i\} \ .
\end{equation}
In terms of representation theory, the associated induced representation of $\mathfrak{g}$ turns out
to be reducible and the physical space of states is its irreducible quotient or the factor representation.
For the coinduction, on the other hand, shortening conditions select a subspace of $\Theta^V_{G/P}$
defined by the equations
\begin{equation}\label{short}
\varphi(Q_\alpha U(\mathfrak{g}))(v) = 0 \quad , \ \mathit{for} \
\alpha \in I = I_V \ .
\end{equation}
Here $v$ is the highest weight vector of the dual $V^\ast$ to the finite dimensional representation $V$.
Obviously, the subspace of functions $\varphi \in \Theta^V_{G/P}$ satisfying this property is
$\mathfrak{g}$-invariant
\begin{equation}
(x\varphi)(Q_i A)(v) = \pm\varphi(Q_i Ax)(v) = 0,\
\textit{for all} \  x\in\mathfrak{g} ,\ A\in
U(\mathfrak{g}). \nonumber
\end{equation}
The fact that $(\ref{short})$ is the right notion of shortening follows from duality between induction
and coinduction, \cite{Blattner} Proposition 1.
Let us now consider two BPS representations associated with the representations $V_1$ and
$V_2$ of $\mathfrak{k}$. According to our discussion above, their tensor product is isomorphic to
$\Theta_{G/K}^{V_1 \otimes V_2'}$. By substituting the explicit form of the isomorphism, $\psi =
(\varphi_1\otimes\varphi_2)\circ\Delta$, we see that BPS conditions imply that elements $\psi$ satisfy
\begin{equation}
\psi(Q_\alpha U(\mathfrak{g})) (v_1\otimes v'_2) = \psi(s(Q_\beta) U(\mathfrak{g}))(v_1\otimes v'_2) = 0,
\end{equation}
for all $\alpha \in I_1$ and $\beta\in I_2$. Note that $s(Q_\beta)  = S_\beta$ are generators
of special superconformal transformations. If the superconformal primary fields are scalars, there
is only one vector in the dual of $V_1\otimes V'_2$ and conditions consequently simplify to
\begin{equation}
\psi(Q_\alpha U(\mathfrak{g})) = \psi(s(Q_\beta) U(\mathfrak{g})) = 0.
\end{equation}
Now that we understand how to construct tensor products we can turn our attention to the
space $\Gamma$ of conformal partial waves. The arguments we have given so far imply that
$\Gamma$ consists of all functions
$$ f \in C^\infty(K\backslash G_{(0)}/K, (V_{(12)} \otimes V_{(34)} \otimes\Lambda \mathfrak{g}_{(1)})^B) $$
such that
\begin{equation}  \label{eq:constraints}
\mathcal{R}_{Q_{\alpha_1}} f (v_{12})= 0 \ , \quad \mathcal{R}_{s(Q_{\alpha_2})} f(v_{12}) = 0\ ,  \quad
\mathcal{L}_{Q_{\alpha_3}} f(v_{34}) = 0 \ , \quad \mathcal{L}_{s(Q_{\alpha_4})} f(v_{34}) = 0
\end{equation}
for all $\alpha_i \in I_i$ and $i=1, \dots, 4$, and $v_{12} = v_1\otimes v'_2,\ v_{34}=v_3\otimes v'_4$.
In order to evaluate the fermionic left and
right invariant vector fields, we first extend $f$ to a function on the supergroup $G$. Then the actions of $\mathcal{L}$ and $\mathcal{R}$ are by the invariant vector fields.

\section{Casimir Equations and Solution Theory}

The main goal of this section is to obtain the Casimir equations for
superconformal partial waves. In the previous section we have realized
the latter through a special subspace of functions on the supergroup.
The Casimir equation is the restriction of the supergroup Laplacian
to this subspace. As we shall show, this equation takes a very simple
form for superconformal groups of type I. The type I condition will
be introduced in the second subsection before we discuss the associated
Casimir equation in the third. The latter turns out to take the form
of a Schroedinger problem that can be solved exactly starting from
spinning bosonic partial waves by a nilpotent perturbation.

\subsection{Casimir equations and Calogero-Sutherland Models}

Our main result in the previous section was to realize the space of
superconformal partial waves as a space of functions on $G_{(0)}$.
The relevant subspace consists of functions that obey a number of
differential conditions involving left and right invariant vector
fields. If all external fields are in long multiplets, we only
need to impose conditions on the left and right invariant vectors
fields that arise from left and right transformations with elements
of the subgroup $K \subset G$. Additional shortening conditions set
some of the fermionic vector fields to zero, as described in
subsection \ref{shortening}.

The Laplacian on a group or a supergroup famously commutes with left
and right invariant vector fields. Hence the constraints we use
to characterize our space $\Gamma$ of superconformal partial waves
respect the action of the Laplacian, i.e.\ the Laplacian descends
to a second order differential operator on $\Gamma$. In principle
it is rather straightforward to work out the Laplacian and its
restriction to $\Gamma$ explicitly. In order to do so, one picks
some coordinates on $G$ and computes e.g.\ the right invariant
vector fields from the Maurer-Cartan form $dg g^{-1}$. Left invariant
vector fields are computed from $g^{-1} dg$. As a result one obtains
two sets of first order differential operators $\mathcal{R}_x$ and
$\mathcal{L}_x$ that involve differentiation with respect to bosonic
and fermionic coordinates as well as multiplication with functions on
$G_{(0)}$ and the products of fermionic coordinates. Differentiation
and multiplication with respect to fermionic coordinates possess a
natural interpretation in terms of (annihilation and creation)
operators on the exterior algebra $\Lambda \mathfrak{g}_{(1)}$. Hence
the left and right actions of the superalgebra $\mathfrak{g}$ on the
space \eqref{eq:GammaVG} of functions in the bosonic group $G_{(0)}$
is given by two sets of first order differential operators in the
bosonic coordinates that take values in the linear maps on the
exterior algebra (the action on $V$ is trivial).

Given such expressions for the left and right invariant vector fields
we can compute the Casimir elements to obtain matrix valued
higher order differential operators that act on the bosonic coordinates
and descend to the space \eqref{eq:CPWfunctions} of superconformal
partial waves. In particular, the quadratic Casimir gives rise to
matrix valued Casimir equations that are second order in derivatives
with respect to the two bosonic cross ratios, i.e.\ the coordinates on
$K\backslash G_{(0)}/K$. According to eqs.\ \eqref{eq:CPWlong}, the
number of matrix components is given by the dimension $M$ that we
introduced in eq.\ \eqref{eq:M}. Note that our formula for $M$ does
not account for shortening conditions and conservation laws. These
can remove some components. We shall discuss this in some examples below.

After an appropriate gauge transformation, the Casimir equations for the
ordinary bosonic partial waves were observed to coincide with the eigenvalue
equations for an integrable two-particle 1-dimensional Schroedinger problem
of Calogero-Sutherland type. Conformal partial waves for correlators of
four scalar fields are wave functions with a potential that contains six
terms each of which is associated with a positive root of an non-reduced
$\textit{BC}_2$ root system. The coupling constants of these six terms, which
describe an external potential and interaction terms for the two particles,
depend on the conformal weights and the dimension $d$ of space-time. If
some of the external fields carry spin, the potential becomes matrix
valued. A general formalism and a few examples were worked out in
\cite{Schomerus:2016epl,Schomerus:2017eny}.

The Casimir equations for superconformal partial waves give rise to certain
supersymmetric Calogero-Sutherland models. Our goal now is to derive these
for a certain subset of superconformal algebras for which the equations take
a particularly simple form. Moreover, they can be solved very systematically.

\subsection{Superconformal groups of type I}

In the following we will focus our analysis of Casimir equations and their solutions to the cases in which the superconformal algebra $\mathfrak{g}$ is of {\it type I}
(in Kac's original notation \cite{Kac:1977em} called {\it case II}). In addition to requiring
that the even part $\mathfrak{g}_{(0)}$ of $\mathfrak{g}$ contains the conformal Lie
algebra $\mathfrak{g}_c=\mathfrak{so}(1,d+1)$ as a direct summand, this means that the
odd subspace decomposes as a direct sum of two irreducible representations of
$\mathfrak{g}_{(0)}$
\begin{equation}
\mathfrak{g}_{(1)} = \mathfrak{g}_+ \oplus \mathfrak{g}_-. \nonumber
\end{equation}
The two modules $\mathfrak{g}_\pm$ are dual to each other. Finally, the action of
$\mathfrak{so}(d)\subset\mathfrak{g}_c$ on the odd subalgebra should decompose into a direct sum of spinor
representations only.

The conditions we listed imply that generators of $\mathfrak{g}_+$ anti-commute with each other and the same is true for $\mathfrak{g}_-$,
$$ \{\mathfrak{g}_\pm , \mathfrak{g}_\pm\} = 0 \ . $$
In addition, the bosonic algebra has the form
\begin{equation}
\mathfrak{g}_{(0)} = [\mathfrak{g}_{(0)},\mathfrak{g}_{(0)}] \oplus \mathfrak{u}(1). \nonumber
\end{equation}
The $\mathfrak{u}(1)$ summand is part of the $R$-symmetry. With respect to its generator,
all elements in $\mathfrak{g}_+$ possess the same $R$-charge. The same is true for the
elements of $\mathfrak{g}_-$, but the $R$-charge of these elements has the opposite value.
Elements in the even subalgebra $\mathfrak{g}_{(0)}$, on the other hand, commute with the
generator of the $\mathfrak{u}(1)$ and hence possess vanishing $R$-charge under this part
the $R$-symmetry.

In superconformal algebras we had seen a different split of the odd generators in section 2, namely the split onto super-translations $Q_\alpha$ and special superconformal transformations $S_\alpha$. This split is not the same as the split into $\mathfrak{g}_\pm$. The anti-commutator of super-translations, for example, gives translations and hence is non-zero in general. Let us denote the intersections of the subspaces $\mathfrak{q}$ and $\mathfrak{s}$ with $\mathfrak{g}_\pm$
by
\begin{equation} \label{eq:split}
\mathfrak{q}_\pm = \mathfrak{q}\cap\mathfrak{g}_{\pm} \quad,\ \mathfrak{s}_\pm =
\mathfrak{s}\cap\mathfrak{g}_\pm.
\end{equation}
The subspaces $\mathfrak{q}_\pm$ and $\mathfrak{s}_\pm$ do not carry a representation of
$\mathfrak{g}_{(0)}$ any more but they do carry a representation of $\mathfrak{k}$. This also
means that in type I superconformal algebras, the action of $\mathfrak{k}$ on super-translations
decomposes into two or more irreducible representations. It turns out that
$$
\text{dim}(\mathfrak{q}_\pm) = \text{dim}(\mathfrak{s}_\pm)  = \text{dim}(\mathfrak{g}_{(1)})/4\ .
$$
It is not difficult to list the superconformal algebras, using e.q. \cite{Kac:1977em}. In $d=4$
dimensions, the superconformal algebras $\mathfrak{sl}(4|\mathcal{N})$ with any number
$\mathcal{N}$ of supersymmetries are of type I. In addition, the type I condition is satisfied
by $\mathfrak{psl}(4|4)$ as well as the series $\mathfrak{sl}(2|\mathcal{N})$ of super-%
conformal algebras in $d=1$ dimension. Finally, it also holds for $\mathfrak{psl}(2|2)$ and for
$\mathfrak{osp}(2|4)$. While the former is also relevant for $d=1$, the latter appears as the
superconformal algebra for systems in $d=3$ with $\mathcal{N} = 2$ supersymmetries. Finally, our
analysis also applies for $d=2$ dimensional systems whose symmetry is a product of type I
superconformal algebras $\mathfrak{sl}(2|\mathcal{N})$ or $\mathfrak{psl}(2|2)$. The discussion
here was restricted to complex Lie superalgebras - for different spacetime signatures one
considers their various real forms.

\subsection{Reduction to the bosonic case}

The advantage of dealing with a type I superalgebra is that there exists a convenient choice
of coordinates on the corresponding supergroup in which the Laplacian takes a simple form.
In the reduction from $G$ to the double coset $K\backslash G/ K$ it is advantageous to work
with coordinates that are based on a $KAK$ decomposition of group elements since the space
$\Gamma$ of partial waves is characterized by how left and right invariant vector fields
$\mathcal{L}_x$ and $\mathcal{R}_x$ act for $x \in K$. When $G$ is a supergroup, the middle
factor $A = A_F$ in such a decomposition would contains the two bosonic coordinates that
parametrize the double coset $K \backslash G_{(0)}/K$ along with all fermionic coordinates
of the supergroup.

In working out the Casimir equations for type I superconformal algebras we shall slightly
deviate from such a parametrization for two reasons. First, as we have stressed before, it
will turn out that the Casimir operator can be written as a sum of the Casimir operator for
bosonic spinning partial waves and a purely fermionic perturbation. In arriving at such a
result, it is a good strategy to keep the bosonic generators of the superconformal algebra
together and not to separate them through some fermionic ones as in the $KA_FK$ decomposition
where exponentials of fermionic generators appear in $A_F$, separating the two bosonic factors
$K$ to the left and right.

In addition, our discussion of shortening conditions shows that, in the case some of the external
operators belong to BPS multiplets, the $K$-covariance conditions that select partial waves
from general functions of the superconformal group are supplemented by similar fermionic
constraints \eqref{eq:constraints}. This makes it natural to choose coordinates in which
such fermionic constraints are as easy to implement as those for the left and right invariant
vector fields associated with the bosonic subalgebra $\mathfrak{k}$. Since we have to
implement constraints on left and right invariant vector fields associated to
the supercharges $Q_\alpha$ it seems natural to separate the supercharges into two
subsets, one to be placed to the left, one to the right and similarly for the fermionic
generators $S_\alpha$.

In type I superconformal algebras there is a natural way to split both $\mathfrak{q}$ and
$\mathfrak{s}$ as described in eq.\ \eqref{eq:split}. All this suggest to introduce the
following parametrization of the superconformal group $G$,
\begin{equation}
G = e^{\bar\sigma^a \bar Y_a} \, \left( K  A_{(0)}  K )\right)\,
e^{\sigma^a Y_a}\ . \label{param}
\end{equation}
Here $A_{(0)}$ is the purely bosonic factor from the $KAK = KA_{(0)} K$ decomposition of
$G_{(0)}$. The elements $\bar Y_a$ form a basis of $\mathfrak{g}_-$ with the index $a$
running through $a = 1, \dots, \text{dim} \mathfrak{g}_-$. Elements of the dual basis in $\mathfrak{g}_+$ are denoted by $Y_a$. When written in terms of the usual supercharges
and special superconformal transformations, the exponents read
$$ \bar\sigma^a \bar Y_a = \bar\sigma^\beta_q Q^-_\beta + \bar \sigma_s^\beta S^-_\beta\quad
, \quad  \sigma^a Y_a = \sigma^\beta_q Q^+_\beta + \sigma_s^\beta S^+_\beta\ . $$
Here $Q^\pm_\beta$ is a basis of $\mathfrak{q}_\pm$ and $S^\pm_\beta$ is a basis of
$\mathfrak{s}_\pm$  so that $\beta$ runs through $\beta = 1, \dots, \text{dim}
\mathfrak{g}_{(1)}/4$. Since we have moved the fermionic generators from $\mathfrak{g}_-$
into the leftmost factor, left translations with elements $x \in \mathfrak{k}$ act on the
corresponding Grassmann coordinates. Hence, the constraints
we impose on functions $f$ when we descent from $G$ to the space of partial waves
determine the combined transformation behavior of bosonic and fermionic coordinates.
When we strip off the fermionic coordinates and describe a function $f$ on the
supergroup in terms of a set of bosonic coefficients, we need to keep the non-trivial
transformation of fermions into account. In other words, we conclude that the space of
superconformal partial waves of a type I superconformal algebra takes the form
\begin{equation}
\Gamma = \Gamma_{K\backslash G_{(0)}/ K}^{V_L ,V_R}  \quad \textit{where} \quad
V_L = V_{(12)} \otimes \Lambda\mathfrak{g}_-^\ast\ , \quad
V_R = V_{(34)}\otimes\Lambda\mathfrak{g}_+^\ast .
\nonumber
\end{equation}
In comparison with our general result \eqref{eq:CPWfunctions}, we have split the
exterior algebra $\Lambda \mathfrak{g}_{(1)} = \Lambda \mathfrak{g}_+ \otimes
\Lambda \mathfrak{g}_-$ and then distributed the two factors such that one is
combined with the left space $V_{(12)}$ and the other with the right space
$V_{(34)}$, whereas before we combined $\Lambda \mathfrak{g}_{(1)}$ with the
left factor. Of course, as a linear space, $\Gamma$ does not depend on such
choices as we remarked above. But as we shall see in a moment, the choice of
$V_L$ and $V_R$ we have made here does allow us to read off the precise spin
content of the unperturbed Hamiltonian $H_0$.
\medskip

We have tried to motivate the choice of coordinates \eqref{param} above, mostly
by anticipating that for this special choice the Laplacian on the superconformal
group is known to assume a remarkably simple form \cite{Schomerus:2005bf,
	Saleur:2006tf,Gotz:2006qp,Quella:2007hr}, namely
\begin{equation}
\Delta = \Delta_0 - 2 D^{ab}\bar \partial_a \partial_b.  \label{eq:main}
\end{equation}
Here, $\Delta_0$ is the Laplacian for a collection of spinning partial waves on
the bosonic conformal group. Such spinning partial waves are characterized by the 
choice of two representations of $K$. In the case at hand, these are the representations 
on the finite dimensional vector spaces $V_L$ and $V_R$ we have introduced above. For a 
superconformal group this Casimir of the bosonic subgroup receives a very simple 
correction: a term that involves only second order derivatives of fermionic
coordinates with bosonic coefficients. In eq.\ \eqref{eq:main} we denoted derivatives with
respect to the Grassmann coordinates $\bar \sigma^a$ by $\bar \partial_a$. Similarly,
derivatives with respect to the coordinates $\sigma^a$ are denoted by $\partial_a$.
The coefficients $D^{ab}$ are matrix elements of the representation of the bosonic
group $G_{(0)}$ on the space $\mathfrak{g}_+$, restricted to the section
$A = A_{(0)} \subset G_{(0)}$ of the bosonic conformal group. Since the representations
of $G_{(0)}$ on $\mathfrak{g}_+$ and on $\mathfrak{g}_-$ are conjugate to each other,
this is equivalent to saying that $-D^{ba}$ are matrix elements for the action of $G_{(0)}$
on $\mathfrak{g}_-$. Either way, the coefficients are functions of the two cross ratios that
are very easy to work out explicitly. We shall see an example below. 

When we choose coordinates through the $KAK$ decomposition of the conformal group, 
the Laplacian was shown to take the form of a matrix Calogero-Sutherland model 
\cite{Schomerus:2016epl,Schomerus:2017eny}. The calculations from these papers can
be copied more or less line by line to derive our eq.\ \eqref{eq:Casimir} for the 
Calogero-Sutherland Hamiltonian $H$ of the superconformal system from our formula 
\eqref{eq:main} for the Laplacian. The only step we briefly want to comment on 
appears at the very end: After restricting the Casimir to the double coset, one 
still needs to conjugate with the square root of the volume of the $K\times K$ 
orbits in $G$ in order to trivialize the measure on the section $A$. The Haar 
measure on the supergroup is known, see e.g. \cite{Quella:2007hr}, and it differs 
from the one on the bosonic conformal group by an additional factor. One may 
observe, however, that this additional factor does neither depend on the bosonic 
cross ratios, nor on the fermionic variables. Hence, the volume of the $K \times K$ 
orbits has the same dependence on the cross ratios as in the bosonic case and it is 
independent of the fermionic variables so that conjugation with the square root of 
the volume turns the restriction of $\Delta_0$ into $H_0$, the same Hamiltonian as 
in the bosonic case, and it leaves the nilpotent potential term $A$ unaffected. The 
final result is indeed given by eq.\ \eqref{eq:Casimir}. The simplicity of this 
Hamiltonian is remarkable and it will allow us to construct solutions of the 
superconformal Casimir equations explicitly from known expressions of bosonic 
spinning partial waves.

Before we explain how this is achieved, we want to close this subsection with a few comments on
multiplet shortening. As we explained above, the shortening conditions \eqref{eq:constraints}
involve expressions for fermionic left and right invariant vector fields. The latter also take
a rather simple form in our coordinates \eqref{param}. Right invariant derivatives with respect
to elements in $\mathfrak{g}_-$ and left invariant derivatives with respect to elements in
$\mathfrak{g}_+$ are particularly simple, namely
\begin{equation}
\mathcal{R}_{\bar Y_a} = \bar \partial_a \quad \textit{and} \quad \mathcal{L}_{Y_a} = -\partial_a\  .
\end{equation}
For the remaining derivatives, things become a bit more difficult, but explicit formulas are
also known, \cite{Quella:2007hr}. For example
\begin{equation}
\mathcal{R}_{Y_a} = -D^b_a \partial_b - \kappa_{ij} D(x^i)^b_a\bar \sigma_b
\mathcal{R}^B_{x^j} - \frac12\kappa_{ij} D(x^i)^c_a D(x^j)^d_b \bar\sigma_c\bar\sigma_d\bar\partial^b
\end{equation}
and similarly for left invariant derivatives with respect to $\bar Y_a$. Here, the indices $i,j$
run over a basis $x^i$ of the even subspace $\mathfrak{g}_{(0)}$, $\kappa$ is the corresponding
Killing form and $\mathcal{R}^B$ are right invariant vector fields on the bosonic subgroup $G_{(0)}$.
We see that shortening conditions are very easy to implement when the subspaces $\mathfrak{Q}_1$ and
$\mathfrak{Q}_2$ we introduced in eq.\ \eqref{spaces} are contained in $\mathfrak{q}_-$ while $\mathfrak{Q}_3$
and $\mathfrak{Q}_4$ are contained in $\mathfrak{q}_+$. In particular, if we have half-BPS
multiplets with $\mathfrak{Q}_1 = \mathfrak{Q}_2 = \mathfrak{q}_-$ and $\mathfrak{Q}_3 =
\mathfrak{Q}_4 = \mathfrak{q}_+$ the superconformal Casimir operator \eqref{eq:main} coincides
with the bosonic $\Delta_0$ and hence superconformal partial waves coincide with bosonic ones.
This applies to the case in which we insert chiral fields in position $1,2$ and anti-chiral
fields in position $3,4$. If, on the other hand, we insert a pair of chiral
and anti-chiral fields in position $1,2$ and $3,4$, the shortening conditions are not quite
as simple. We shall study these in more detail for the example we consider in the subsequent
section.

\subsection{Nilpotent perturbation theory}

Having seen that for type I superconformal symmetry the Casimir equation can be regarded as a
nilpotent perturbation of Casimir equations for a set of spinning bosonic conformal partial waves,
our strategy is to construct supersymmetric partial waves as a perturbation of spinning partial waves.
Since the perturbing term $A$ is nilpotent, we can obtain exact formulae at some finite order, which
depends on the precise setup, such as the dimension $d$ and possible shortening conditions for some
or all of the external operators. For $\mathcal{N}=2$ supersymmetry in $d=1$ and also for the kind
of mixed correlators in $\mathcal{N}=1$ superconformal theories in $d=4$ we intend to study in
our upcoming work, the second order perturbation turns out to be exact. The general methods to
solve for eigenfunctions of a Hamiltonian $H= H_0 + A$ in terms of those of $H_0$ are certainly
well established. In our exposition we shall follow Messiah, \cite{messiah1962quantum}. For presentational
purposes we shall assume that $H$ and $H_0$ have discrete spectra and finite dimensional
eigenspaces. By a limiting process, the construction can be extended to more general spectra.

Let us first set up a bit of notation. The Hilbert space on which the operators act is denoted by $\mathcal{H}$ and $H_0$ is assumed to be hermitian. We shall denote by $V^0_n$ and $V_n$
the eigenspaces of $H_0$ and $H$ with eigenvalues $\varepsilon^0_n$ and $\varepsilon_n$. Furthermore,
let us denote the projectors to these eigenspaces by $P^0_n$ and $P_n$,
respectively. Next we introduce the two resolvents
\begin{equation}
G_{[0]}:\mathbb{C}\xrightarrow{}L(\mathcal{H})\ ,\quad  G_{[0]}(z) = (z-H_{[0]})^{-1} \  \nonumber
\end{equation}
for the Hamilton operators of the unperturbed and the perturbed system. Here, the subscript $[0]$ means
that we either put zero or not, i.e.\ $G_{[0]}$ is a shorthand for the pair $G_0, G$ etc. respectively. Obviously, these resolvents can be expanded in the projectors $P_n^{[0]}$ with simple poles at the
eigenvalues $\varepsilon_n^{[0]}$ of $H_{[0]}$,
\begin{equation}
G_{[0]}(z) = \sum_n \frac{1}{z-\varepsilon^{[0]}_n} P^{[0]}_n. \nonumber
\end{equation}
Conversely, the projectors $P_n^{[0]}$ are obtained as the residues of the resolvents,
\begin{equation}
P^{[0]}_n = \frac{1}{2\pi i}\oint_{\Gamma_n} G_{[0]}(z)dz, \nonumber
\end{equation}
where $\Gamma_n$ is a small contour encircling $z=\varepsilon^{[0]}_n$ and none of the other eigenvalues.

Inserting the relation $H = H_0 + A$ between the two Hamilton operators into the resolvent $G$, one can
perform an expansion in $A$ to obtain
\begin{equation}
G = G_0 \sum_{n=0}^{\infty} (A G_0)^n = G_0 \sum_{n=0}^{N} (A G_0)^n. \nonumber
\end{equation}
Here we have truncated the sum to a finite order $N$ in view of our application to a nilpotent perturbation
$A$ of order $N$. Note that $A^N = 0$ implies $(AG_0)^N=0$ in our application since $G_0$ acts diagonally on
$\mathcal{H} = L^2(\mathbb{C}^m)\otimes\mathbb{C}^l$ and $A$ is a triangular matrix of functions.
Computing residues of the previous expansion for $G$ at $\varepsilon^0_i$ we obtain
\begin{equation}
P_i = P^0_i + \sum_{n=1}^{N} \text{Res}(G_0 (AG_0)^n,\varepsilon^0_i)\equiv P^0_i + P^{(1)}_i + ...
+ P^{(N)}_i. \label{proj}
\end{equation}
with
\begin{eqnarray}
P^{(1)}_i & = & P^0_i A S_i + S_i A P^0_i , \label{eq:P1}\\[2mm]
P^{(2)}_i & = & P^0_i A S_i A S_i + S_i A P^0_i A S_i + S_i A S_i A P^0_i \nonumber \\[2mm]
& & \quad -
P^0_i A P^0_i A S_i^2 - P^0_i A S_i^2 A P^0_i - S_i^2 A P^0_i A P^0_i,\label{eq:P2}
\end{eqnarray}
and so on. Here, the symbol $S_i$ denotes the following operator
\begin{equation}
S_i = \sum_{j\neq i}\frac{P^0_j}{\varepsilon^0_i - \varepsilon^0_j}. \label{eq:S}
\end{equation}
Since the sum over $j$ is restricted to $j \neq i$ we infer that $S_i P^0_i = P^0_i S_i = 0$,
a property we shall use frequently below.

The idea now is to solve the eigenvalue equation $H |\psi\rangle = \varepsilon |\psi\rangle$
by transporting the operator identity $H P_i = \varepsilon_i P_i$ to the eigenspace $V_i^0$ of the
unperturbed Hamiltonian. We can do so if the restriction of the projectors $P_i:V^0_i
\xrightarrow{}V_i$ and $P^0_i:V_i\xrightarrow{}V^0_i$ are vector space isomorphisms. If that
is the case, we use $P_i^0$ to transport the two sides of $HP_i = \varepsilon P_i$ to
$V_i^0$, i.e. we define
\begin{align*}
& K_i = P^0_i P_i P^0_i = P^0_i - P^0_i A S_i^2 A P^0_i + \dots  \quad \textit {and} \\[2mm]
& H_i = P^0_i H P_i P^0_i = \varepsilon^0_i K_i + P^0_i A P^0_i + P^0_i A S_i A P^0_i+\dots.
\end{align*}
In the evaluation of $K_i$ we have inserted the expansion \eqref{proj} of $P_i$ and made use
of the fact that $P_i^{0} S_i = S_i P_i^0 = 0$. We have only displayed the terms up to second
order in $A$. To evaluate $H_i$ we inserted $H = H_0 + A$ and used that $P_i^0 H_0= \varepsilon_i^0
P_i^0$ to write the first term as $\varepsilon_i^0K_i$. In the second term we inserted once again
the expansion \eqref{proj} of $P_i$, but this time only to first order since $P_i$ is multiplied
by $A$ already. With these notations, the original eigenvalue equation becomes
\begin{equation}
H_i |\psi\rangle = \varepsilon_i K_i |\psi\rangle \label{gen}
\end{equation}
for $|\psi\rangle \in V_i^0$. Once we have found a solution to this equation, we can map it
back to $V_i$ using $P_i$ to find the eigenvectors $P_i |\psi\rangle$ of $H$ in $V_i$.

\section{An Example: $\mathcal{N}=2$ Supersymmetry in $d=1$}

The goal of this section is to illustrate the general theory we have developed in the
previous sections at the example of $\mathcal{N}=2$ supersymmetry in $d=1$ dimensions,
i.e.\ for the conformal superalgebra $\mathfrak{su}(1,1|1)$. We will first provide all
the necessary mathematical background by describing the algebra itself and its various
subalgebras and the Casimir elements. Then we spell out the Casimir equation as a
nilpotent perturbation of a certain set of Calogero-Sutherland models. The nilpotent
terms are of order two and we will explicitly construct the exact solution in the
third subsection. These are mapped to blocks for long multiplets that were originally
constructed in \cite{Cornagliotto:2017dup} in section 4.4. Finally, we also discuss
shortening conditions and identify the relevant blocks.

\subsection{The Lie superalgebra and representation theory}

The $\mathcal{N}=2$ superconformal algebra in $d=1$ dimension is the Lie superalgebra
$\mathfrak{su}(1,1|1)$, a real form of $\mathfrak{g}=\mathfrak{sl}(2|1)$. We will
work with the complexification $\mathfrak{sl}(2|1)$ until the very end when we choose
an appropriate real slice of the obtained integrable model. In the notation introduced
in the previous sections, $\mathfrak{g}=\mathfrak{g}_{(0)}\oplus\mathfrak{g}_{(1)}$. The bosonic
subalgebra $\mathfrak{g}_{(0)}$ is generated by the dilation $D$, a single translation $P$
along with a corresponding special conformal transformations $K$, and an $R$-charge $R$.
The fermionic subspace, on the other hand, contains four basis elements which we shall denote
as $Q_\pm$ and $S_\pm$. The lower indices of fermionic generators denote the eigenvalues
of $\text{ad}_{R}$ and are degrees of elements in a $\mathbb{Z}$-grading. In this case
the subalgebra $\mathfrak{k}$ is spanned by the generators of dilations $D$ along with
the $R$-charge $R$. The non-trivial (anti-)commutation relations read
\begin{align*}
& [D,P]=P\ ,\quad [D,K]=-K \ ,\quad \ [K,P]=2D\ , \quad [D,R]=[P,R]=[K,R]=0,\\[2mm]
& [K,Q_{\pm}]=S_{\pm}\ ,\quad  [P,S_{\pm}]=-Q_{\pm}\ ,\quad [D,Q_{\pm}]=\frac12Q_\pm,\
[D,S_\pm] = -\frac12 S_\pm,\\[2mm]
&\{Q_\pm,Q_\pm\}=\{S_\pm,S_\pm\}=\{Q_\pm,S_\pm\}=0\ ,\quad \{Q_+, Q_-\}=P\  ,\\[2mm]
& \{S_+,S_-\}=K\ ,\quad \{S_\pm,Q_\mp\}=D\pm\frac12 R\ .
\end{align*}
The even part $\mathfrak{g}_{(0)}$ is the usual one dimensional conformal Lie algebra
$\mathfrak{su}(1,1) = \mathfrak{so}(1,2)$, extended by a $U(1)$ $R$-symmetry, that is
\begin{equation}
\mathfrak{g}_{(0)}=\mathfrak{so}(1,2)\oplus \mathfrak{u}(1).  \nonumber
\end{equation}
Representations $[j,q]$ of this bosonic Lie algebra are labeled by a spin $j$ and an
$R$-charge $q$. For finite dimensional (non-unitary) representations, $j$ is
half-integer while $q$ can be any complex number. We see that the odd subspace
$\mathfrak{g}_{(1)}$ decomposes into a sum of two irreducible representations,
\begin{align*}
& \mathfrak{g}_{(1)} = \mathfrak{g}_+ \oplus \mathfrak{g}_- = [1/2,1]\oplus [1/2,-1].
\end{align*}
Since the subalgebra $\mathfrak{k}$ is generated by two $U(1)$ generators, the
dilation $D$ and the $R$-charge $R$, all its irreducible representations
$(\Delta,q)$ are one-dimensional. When we restrict the irreducible representations
$\mathfrak{g}_\pm$ to $\mathfrak{k}$, they decompose into a sum of two irreducibles
each,
\begin{equation}
\mathfrak{g}_+ = \mathfrak{q}_+ \oplus \mathfrak{s}_+ \quad , \quad
\mathfrak{g}_- = \mathfrak{q}_- \oplus \mathfrak{s}_-\ ,
\end{equation}
where
\begin{equation}
\mathfrak{q}_\pm = (1/2,\pm 1) \quad , \quad \mathfrak{s}_\pm = (-1/2,\pm 1) \ .
\end{equation}
Recall that $\mathfrak{q}_\pm$ are the spaces spanned by $Q_\pm$, respectively, and
the same for $\mathfrak{s}_\pm$. In our analysis of the Casimir equations for generic
long multiplets it is also important to know the representation content of $\Lambda
\mathfrak{g}_\pm$ which is given by
\begin{equation}
\Lambda\mathfrak{g}_\pm = (0,0) \oplus (1/2,\pm 1) \oplus (-1/2,\pm 1) \oplus (0,\pm 2)\ .
\end{equation}
Let us finally recall that the superalgebra $\mathfrak{sl}(2|1)$ possesses two algebraically
independent Casimir elements, one of second order and one of third. The quadratic Casimir
element is given by
\begin{equation}
C_2=-D^2+\frac14 R^2 + \frac12 \{K,P\} +\frac12 [Q_+,S_-] + \frac12 [Q_-,S_+]. \label{e1}
\end{equation}
The cubic Casimir element, on the other hand, reads
\begin{align*}
C_3 & = \big(D^2-\frac14 R^2-PK\big)R - Q_+ S_- \big(D + \frac32 R\big) + Q_-
S_+ \big(D-\frac32 R\big)\\[2mm]
& + K Q_+ Q_- - P S_-S_+ - D - \frac12 R.
\end{align*}
Long multiplets of the superalgebra $\mathfrak{sl}(2|1)$ can be distinguished by the values
of these two Casimir elements. For short multiplets this is not the case. For those multiplets
both Casimirs are zero regardless of the precise representation we consider, see e.g.\ \cite{Scheunert:1976wj,Cordova:2016emh}.

\subsection{The supergroup and Hamiltonian reduction}

According to the general prescription, the coordinates on the supergroup $G=SU(1,1|1)$ are introduced
by writing its "elements" as
\begin{equation}
g = e^{\bar\varrho Q_- + \bar\sigma S_-} e^{\kappa R} e^{\nu_1 D} e^{\mu (P+K)} e^{\nu_2 D}
e^{\sigma S_+ + \varrho Q_+}. \label{e2}
\end{equation}
Two such elements can be formally multiplied together using the Campbell-Baker-Hausdorff
formula to obtain an element of the same form. In the following, we will treat eq.\ (\ref{e2})
as an honest decomposition, like in ordinary Lie theory, bearing in mind that all our constructions
are rigorously formulated in terms of the structure algebra $\mathcal{A}(G)$. We can now execute the
steps we described in the first section 3.1 to find Laplacian and descend to the double coset
$K\backslash G_{(0)} /K$.

Since the algebra $\mathfrak{k}$ is abelian, the spaces $V_i$ and hence also $V_{(12)}$ and
$V_{(34)}$ are all one-dimensional. We will assume that the $R$-charges $q_i, i=1, \dots,4$
of the four external fields sum up to $\sum q_i = 0$. Recall that the Laplacian acts on a
space of functions \eqref{eq:CPWfunctions} that take values in $B$-invariants where in the
case at hand, $B$ coincides with the $R$-symmetry group $U(1)$ and hence it is generated by
a single element $R$. With our assumption of vanishing total $R$-charge of the external
superprimaries we find
\begin{equation}
\left(\Lambda \mathfrak{g}_{(1)}\otimes V_{(12)} \otimes V_{(34)}\right)^B
= \text{span}\{1,\bar\sigma \sigma, \bar\sigma \varrho,\bar\varrho \sigma,\bar\varrho
\varrho,\bar\sigma \bar\varrho\sigma \varrho\}\ .
\end{equation}
Each function on the one-dimensional coset space $K \backslash G_{(0)} /K$ that takes values
in this subspace can extended to a covariant function $f$ on the entire supergroup as
\begin{align*}
f&(\mu,\kappa,\nu,\sigma,\varrho)  =
e^{a\nu_1 + b\nu_2 + q\kappa} f_1 +  \\[2mm]
& + e^{(a+\frac12)\nu_1 + (b-\frac12)\nu_2 + (q+1)\kappa}f_{2}\, \bar\sigma\sigma +
e^{(a+\frac12)\nu_1 + (b+\frac12)\nu_2 + (q+1)\kappa}f_{3} \, \bar\sigma\varrho +\\[2mm]
& + e^{(a-\frac12)\nu_1 +  (b-\frac12)\nu_2 + (q+1)\kappa}f_{4}\, \bar\varrho\sigma +
e^{(a-\frac12)\nu_1 + (b+\frac12)\nu_2 + (q+1)\kappa}f_{5}\, \bar\varrho\varrho +\\[2mm]
& + e^{a\nu_1 + b\nu_2 + (q+2)\kappa} f_6\, \bar\sigma\bar\varrho\sigma\varrho,
\end{align*}
where the six real component functions $f_1,...,f_6$ depend on the variable $\mu$ that
parametrizes the double coset. The behavior of the individual terms under the left and
right action of $K$ is determined by the parameters $(\Delta_i,q_i)$ of the external
fields. Their values are $a = \Delta_2-\Delta_1$, $b= \Delta_3-\Delta_4$  and $q =
q_1+q_2$. Recall that we assumed that $\sum q_i = 0$ so that $q_1+q_2 = -q_3-q_4$. The
precise form of the $\nu_i$ and $\kappa$-dependent prefactor depends on the fermionic
coordinates they are multiplied with. The first term in the expansion above, one that contains
no fermionic coordinates, is multiplied by the character of $K \times U(1)_D$ on $V_{(12)}
\times V_{(34)}$ where $U(1)_D$ denotes the $U(1)$ subgroup of the right factor $K$ that
is associated with dilation. In the remaining terms, this basic character is multiplied
with the character of $K \times U(1)_D$ on the corresponding product of fermionic variables,
taking into account that $\bar\sigma$ and $\bar\varrho$ have $R$-charge $q=1$ under left
multiplication with elements in $U(1)_R$ while $\sigma$ and $\varrho$ transform trivially.
With respect to left dilations, $\bar \sigma$ and $\bar \rho$ have weight $\Delta = \pm 1/2$.
The weights of $\sigma$ and $\varrho$ are the same but there is an additional minus sign
since in the right regular action group element is inverted. Notice that we are working
in one-dimensional theory and did not assume that it is a holomorphic part of a two-dimensional
theory. This is why our conventions for the relation between the parameters $a,b$ with the
external conformal weights differ from the standard ones by a factor of two, see
\cite{Isachenkov:2017qgn}.
\medskip

The Laplace-Beltrami operator is obtained by substituting explicit expressions for the
left or right invariant vector fields in the quadratic Casimir $(\ref{e1})$. Once we
carry out all derivatives with respect to $\kappa, \nu_i$ and the four fermionic
variables we end up with a second order differential operators that acts on the six
component functions $f_1(\mu), \dots, f_6(\mu)$. The corresponding eigenvalue problem assumes the form of a matrix Schroedinger equation
\begin{equation}
H f  = \lambda f, \label{lap}
\end{equation}
where the Hamilton operator is of the form $H = H_0 + A$ with
\begin{align*}  \label{eq:H0}
H_0 = & \, \text{diag} ( H_\text{PT}^{(a,b)}-\frac{(q-1)^2}{4} , H_\text{PT}^{(a+\frac12,b-\frac12)}-
\frac{q^2}{4} , H_\text{PT}^{(a+\frac12,b+\frac12)}-\frac{q^2}{4},\\[2mm]
& \hspace*{8mm} H_\text{PT}^{(a-\frac12,b-\frac12)}-\frac{q^2}{4},H_\text{PT}^{(a-\frac12,b+\frac12)}-\frac{q^2}{4},
H_\text{PT}^{(a,b)}-\frac{(q+1)^2}{4}),
\end{align*}
and a nilpotent perturbation
\begin{equation}
A=
\begin{pmatrix}
0 & -\sin\mu & \cos\mu & -\cos\mu & -\sin\mu & 0\\
0 & 0 & 0 & 0 &  0 & \sin\mu\\
0 & 0 & 0 & 0 & 0 & -\cos\mu \\
0 & 0 & 0 & 0 & 0 &  \cos\mu\\
0 & 0 & 0 & 0 & 0 & \sin\mu\\
0 & 0 & 0 & 0 & 0 & 0
\end{pmatrix}.\nonumber
\end{equation}
The unperturbed Hamiltonian $H_0$ contains six individual Hamiltonians $H_\text{PT}^{(\alpha,\beta)}$
with a Poeschl-Teller potential,
\begin{equation}
H_\text{PT}^{(\alpha,\beta)} = - \frac{1}{4}\partial_\mu^2  - \frac{\alpha\beta}{\sin^2 \mu} +
\frac{(\alpha+\beta)^2-\frac{1}{4}}{\sin^2 2\mu} \ .
\end{equation}

The eigenvalue equation for these Poeschl-Teller Hamiltonians are Casimir equations for various
four point functions of the component fields with respect to the ordinary bosonic conformal
symmetry. A superfield for an $\mathcal{N}=2$ theory in $d=1$ dimensions contains four
component fields of weight $\Delta_i+n, n=0,1/2,1$ and $R$-charge $q_i + m, m=0,\pm 1$. We use
superconformal symmetry to set the fermionic coordinates in front of half of these component
fields to zero so that we remain with correlators of the scalar superprimaries $\phi$ and its
fermionic superpartners $\psi$ and $\bar \psi$. Only six among them satisfy $R$-charge
conservation and hence are non-zero, namely
\begin{eqnarray*}
	\langle \phi \phi \phi \phi \rangle\quad , \quad \langle \phi \psi \phi \bar \psi \rangle\quad ,
	\quad \langle \phi \psi \bar \psi \phi \rangle \\[2mm]
	\langle \psi \phi \phi \bar \psi \rangle\quad , \quad \langle \psi \phi \bar \psi \phi \rangle\quad ,
	\quad \langle \psi \psi \bar \psi \bar \psi \rangle \ .
\end{eqnarray*}
The precise combination of external weights and charges of the involved components explains the
different values of the coupling constants and ground state energies in the six Poeschl-Teller
Hamiltonians appearing in $H_0$.

\subsection{Solutions on a compact domain}

We will now apply the exact nilpotent perturbation theory we reviewed in section 3.4 to solve the eigenvalue problem $(\ref{lap})$ for the case $a=b=q=0$. It is seen that $A^2$
has a single non-zero entry, namely a constant $-2$ in the top right corner, and $A^3=0$.
Hence our perturbative expansion truncates after the second order. For simplicity we shall
first solve the eigenvalue problem in the trigonometric case in which the potential
diverges for $\mu = 0,\pi/2$. Hence, the spectrum of the Schroedinger problem on the
interval $\mu\in[0,\pi/2]$ is discrete.

According to our general discussion, we first need to spell out the solution of the
unperturbed problem, i.e.\ provide the eigenfunctions of the Poeschl-Teller Hamiltonians
that appear along diagonal of $H_0$. Non-singular eigenfunctions of
$$
H_\text{PT}^{(0,0)}-1/4\ ,\quad  H_\text{PT}^{(1/2,-1/2)} = H_\text{PT}^{(-1/2,1/2)}\ ,
\quad H_\text{PT}^{(1/2,1/2)}= H_\text{PT}^{(-1/2,-1/2)}
$$
will be denoted by $\psi_n,\phi_n$ and $\chi_n$ with $n = 0,1, \dots $ integer, respectively.
Explicitly, they are given by
\begin{align*}
& \psi_n = a_n\sin^{1/2}\mu \cos^{1/2}\mu\ P_n(1-2\sin^2\mu)\ ,\quad
\varepsilon^{0}_{0,n} = n(n+1),\\[2mm]
& \phi_n = \frac{1}{n+1} b_n\sin^{3/2}\mu \cos^{1/2}\mu\ P_n^{(1,0)}(1-2\sin^2\mu),\
\varepsilon_{1,n}^0 = (n+1)^2,\\[2mm]
& \chi_n = c_n\sin^{1/2}\mu \cos^{3/2}\mu\ P_n^{(0,1)}(1-2\sin^2\mu),\
\varepsilon_{1,n}^0 = (n+1)^2.
\end{align*}
Here, $P_n^{(\alpha,\beta)}$ denote Jacobi polynomials, $P_n = P_n^{(0,0)}$ are the Legendre
polynomials and the normalisation constants are
\begin{equation}
a_n = \sqrt{2(2n+1)},\ b_n=2(n+1)^{3/2},\ c_n = 2\sqrt{n+1}. \nonumber
\end{equation}
With this choice, each set of wave functions forms an orthonormal basis for the space of functions defined on the interval $[0,\pi/2]$ which vanish on the boundary, with respect to the usual scalar
product,
$$ (g_1, g_2) = \int_0^{\frac{\pi}{2}} d\mu g_1 (\mu) \bar g_2 (\mu) \  $$
for which the Poeschl-Teller Hamiltionians are Hermitian. When we displayed the eigenvalues
$\varepsilon$ of the Poeschl-Teller Hamiltonians we have already introduced the label $ i =
(\sigma,n), \sigma = 0,1$ that enumerates the various eigenspaces of the unperturbed
Hamiltonian $H_0$. We can now also display the associated projectors $P^0_i = P^0_{\sigma,n}
= P^0_{\varepsilon_{\sigma,n}}$. They are given by
\begin{align*}
& P^0_{n(n+1)} f = (\psi_n,f_1)\psi_n e_1 + (\psi_n,f_6)\psi_n e_6,\\[2mm]
& P^0_{(n+1)^2}f = (\phi_n,f_2)\phi_n e_2 + (\phi_n,f_5)\phi_n e_5 +
(\chi_n,f_3)\chi_n e_3 + (\chi_n,f_4)\chi_n e_4, \nonumber
\end{align*}
where $f=(f_1,...,f_6)^T$ is a six component column of functions in $\mu$. In order to find
eigenvectors of $H$ we first need to solve equation $(\ref{gen})$. To do this, we introduce
the following two integrals
\begin{align*}
& I_1(m,n) = \int_0^{\pi/2} d\mu\ \phi_m \psi_n\sin\mu =
\frac{b_m a_n}{8(m+1)}\int_{-1}^{1} dx\ (1-x) P_m^{(1,0)}(x) P_n(x),\\[2mm]
& I_2(m,n) = \int_0^{\pi/2} d\mu\ \chi_m \psi_n\cos\mu =
\frac18 c_m a_n\int_{-1}^1 dx\ (1+x) P_m^{(0,1)}(x) P_n(x),
\end{align*}
where we performed the substitution to a new variable $x=\cos2\mu$ that takes values in $x\in[-1,1]$.
Using the relations
\begin{equation}
(1-x) P_n^{(1,0)} = P_n- P_{n+1}\ , \quad  (1+x) P_n^{(0,1)} = P_n + P_{n+1}, \nonumber
\end{equation}
and orthogonality of Legendre polynomials, we can evaluate $I_1$ and $I_2$ as
\begin{align*}
& I_1(m,n)=\frac{a_n b_m}{4(m+1)(2n+1)}(\delta_{mn}-\delta_{m+1,n}) =
\sqrt{\frac{m+1}{2(2n+1)}}(\delta_{mn}-\delta_{m+1,n}),\\[2mm]
& I_2(m,n) = \frac{a_n c_m}{4(2n+1)}(\delta_{mn}+\delta_{m+1,n}) =
\sqrt{\frac{m+1}{2(2n+1)}}(\delta_{mn}+\delta_{m+1,n}).
\end{align*}
These results imply that
\begin{equation}
P^0_i A P^0_i = 0,\  P^0_i A S_i A P^0_i=0. \nonumber
\end{equation}

Here, the index $i = (\sigma,n)$ runs over $\sigma = 0,1$ and $n = 0,1,2, \dots$ or
alternatively the corresponding set of eigenvalues $n(n+1)$ and $(n+1)^2$. Looking back
at the relations between projectors, we see that any $|\psi\rangle\in V^0_n$ solves the
eigenvalue problem and has the eigenvalue $\varepsilon_i = \varepsilon^0_i$. In particular,
$H_0$ and $H$ have the same spectra. To get the eigenvectors of $H$ all we have to do
is to apply the projectors $P_i$ to $|\psi\rangle$. Using equation \eqref{proj} and
the expressions \eqref{eq:P1} and \eqref{eq:P2} for $P^{(1)}_i, P^{(2)}_i$ we obtain
the following set of linearly independent eigenfunctions of the perturbed Hamiltonian, i.e.\ the Laplacian on the supergroup,
\begin{align*}
& f^{(1)}_n = \psi_n e_1,\\
& f^{(2)}_n = \phi_n e_2 - \frac{1}{\sqrt{2(n+1)(2n+1)}} \psi_n e_1 - \frac{1}{\sqrt{2(n+1)(2n+3)}} \psi_{n+1} e_1,\\
& f^{(3)}_n = \chi_n e_3 + \frac{1}{\sqrt{2(n+1)(2n+1)}} \psi_n e_1 - \frac{1}{\sqrt{2(n+1)(2n+3)}} \psi_{n+1} e_1,\\
& f^{(4)}_n = \chi_n e_4 - \frac{1}{\sqrt{2(n+1)(2n+1)}} \psi_n e_1 + \frac{1}{\sqrt{2(n+1)(2n+3)}} \psi_{n+1} e_1,\\
& f^{(5)}_n = \phi_n e_5 - \frac{1}{\sqrt{2(n+1)(2n+1)}} \psi_n e_1 - \frac{1}{\sqrt{2(n+1)(2n+3)}} \psi_{n+1} e_1,\\
& f^{(6)}_n = \psi_n e_6 + \frac{1}{\sqrt{2n(2n+1)}} (\phi_{n-1}(-e_2-e_5) +\chi_{n-1}(-e_3+e_4)) -\\
& -\frac{1}{\sqrt{2(n+1)(2n+1)}} (\phi_n (e_2+e_5) +\chi_n (-e_3+e_4)) + \frac{2}{n(n+1)} \psi_n e_1.
\end{align*}
Note that the superscript $(k)$ labels different solutions of our matrix Schroedinger
equation. Each of the eigenfunctions $f^{(k)}$ has six components.

We conclude this subsection with a couple of remarks about the obtained set of eigenfunctions. By completeness of eigenfunctions of each Poeschl-Teller Hamiltonian, the eigenfunctions of $H_0$ are also complete in the Hilbert space of physical wave functions. However, the solution $f^{(6)}_n$ is not well-defined for $n=0$ and it is therefore discarded. Indeed, the perturbed Hamiltonian is seen to be no longer diagonalizable on the full Hilbert space, but it is diagonalizable on a codimension-one subspace. Non-diagonalizability is a known feature of the Laplacian on supergroups, \cite{Schomerus:2005bf,Saleur:2006tf}, and is related to the presence of atypical modules in the decomposition of the regular representation. In our case, as mentioned above, atypical (short) representations can appear only for eigenvalue zero, consistent with the findings here. In the conformal field theory language, the number of conformal blocks reduces when the field in the intermediate channel is BPS.

\subsection{Conformal blocks and comparison}

In order to obtain conformal blocks and to compare them with the expressions that were found
in \cite{Cornagliotto:2017dup} we need to perform a few simple steps. First of all, we need to
adapt the solution of the trigonometric model we constructed in the previous subsection to the
hyperbolic theory. This is fairly straightforward. The hyperbolic model has a continuous spectrum
that is parametrized by $\lambda$ instead of the discrete parameter $n$ and the eigenfunctions do
not depend on a variable $\mu$ but rather on $u = 2i\mu$ which takes values in the non-negative
real numbers.

The building blocks of the solution for the hyperbolic models can be found within the
following family of functions
\begin{equation}
\Psi^{(a,b)}_\lambda = \left(\frac{4}{y}\right)^{a+\frac12}(1-y)^{\frac12 a-\frac12 b+\frac14} \ _2F_1\Big(\frac12+a+\lambda,\frac12+a-\lambda,1+a-b,\frac{y-1}{y}\Big),
\end{equation}
where the variable $y$ is related to $u$ as
$$ y = \cosh^{-2} \frac{u}{2} \ . $$
For our special values of the parameters $a,b$ we introduce in particular
\begin{equation}
\Psi_\lambda = (i\lambda)^{1/2} \Psi^{(0,0)}_\lambda\, ,\quad
\Phi_\lambda = \frac12 (i\lambda)^{3/2} \Psi^{(\frac12,-\frac12)}_\lambda\, ,\quad
X_\lambda = \frac12 (i\lambda)^{1/2} \Psi^{(\frac12,\frac12)}_\lambda. \label{eq:functions}
\end{equation}
When these are evaluated at special points $\lambda$, we obtain the polynomial building
blocks of our solution for the trigonometric model, more precisely
\begin{equation}
\Psi_{\lambda=-n-\frac12} = \psi_{n}\, ,\quad
\Phi_{\lambda=-n-1} = \phi_{n},\
X_{\lambda=-n-1} = \chi_{n}.
\end{equation}
With this in mind, it is now easy to write down a set of solutions for the hyperbolic model that, upon continuation to special values of $\lambda$, reproduces our previous solution of
the trigonometric theory. We denote these functions by $F^{(i)}_\lambda$, $i=1,...,6$. They are written explicitly in Appendix A.

Functions $F^{(1)}_{\lambda}, F^{(6)}_{\lambda}$ have eigenvalues $\varepsilon=\lambda^2-\frac14$ for $H$ and $c_3=\mp\varepsilon$
for $C_3$, respectively. The other four functions have eigenvalue $\varepsilon=\lambda^2$ for $H$ and zero for $C_3$. The unitarity bound occurs when atypical representations propagate in the intermediate channel, that is, for $\varepsilon=0$. Thus, we can restrict our attention to the cases $\varepsilon>0$.

The functions $F^{(i)}$ are solutions of the matrix Calogero-Sutherland model that are regular near the wall at $u=0$, i.e.\ they are true physical wave functions of the model in which incoming and outgoing
waves are superposed in a very particular way. There exists a natural decomposition of conformal
partial waves into a block and its shadow which we now mimic for the wave functions $F^{(i)}$, i.e. we
want to decompose these functions into a linear combination of two pieces with purely ingoing and
outcoming behavior at $u = \infty$. For the building blocks $\Psi^{(a,b)}_\lambda$ the relevant
decomposition
\begin{equation}
\Psi^{(a,b)}_\lambda = \Psi^{(a,b)}_{\lambda,+} + \Psi^{(a,b)}_{\lambda,-}, \nonumber
\end{equation}
involves
\begin{equation}
\Psi^{(a,b)}_{\lambda,\pm} =c(\pm\lambda,a,b)  4^{\pm\lambda} (1-y)^{\frac12 a-\frac12 b +\frac14} y^{\mp \lambda}\ _2 F_1\Big(\frac12+a\mp\lambda,\frac12-b\mp\lambda,1\mp2\lambda,y\Big), \nonumber
\end{equation}
with a prefactor $c$ given by
\begin{equation}
c(\lambda,a,b) = 4^{-\lambda+a+\frac12} \frac{\Gamma(a-b+1) \Gamma(2\lambda)}{\Gamma(\frac12+\lambda+a)
	\Gamma(\frac12+\lambda-b)}.  \nonumber
\end{equation}

Thus, the wavefunctions $F^{(i)}$ give two families of solutions which are obtained by expressing $F^{(i)}_\lambda$ in terms of $\Psi^{(a,b)}_\lambda$ and attaching an index $+$ (respectively $-$) to them. For $\lambda,\varepsilon>0$, the set of solutions which decay at infinity is 
\begin{align}\label{solutions}
&  G^{(1)}_{\lambda} = (i\lambda)^{1/2} \Psi^{(0,0)}_{\lambda,-} e_1,\nonumber\\
&  G^{(2)}_{\lambda} = \frac12 (i\lambda)^{3/2} \Psi^{(\frac12,-\frac12)}_{\lambda,-} e_2 +
\sqrt{\frac{i}{4\lambda}} \Psi^{(0,0)}_{\lambda+\frac12,-} e_1 + \sqrt{\frac{i}{4\lambda}}
\Psi^{(0,0)}_{\lambda-\frac12,-} e_1 ,\nonumber\\
&  G^{(3)}_{\lambda} = \frac12 (i\lambda)^{1/2} \Psi^{(\frac12,\frac12)}_{\lambda,-} e_3 -
\sqrt{\frac{i}{4\lambda}} \Psi^{(0,0)}_{\lambda+\frac12,-} e_1 + \sqrt{\frac{i}{4\lambda}}
\Psi^{(0,0)}_{\lambda-\frac12,-} e_1,\nonumber\\
&  G^{(4)}_{\lambda} = \frac12 (i\lambda)^{1/2} \Psi^{(\frac12,\frac12)}_{\lambda,-} e_4 +
\sqrt{\frac{i}{4\lambda}} \Psi^{(0,0)}_{\lambda+\frac12,-} e_1 - \sqrt{\frac{i}{4\lambda}}
\Psi^{(0,0)}_{\lambda-\frac12,-} e_1,\\
&  G^{(5)}_{\lambda} = \frac12 (i\lambda)^{3/2} \Psi^{(\frac12,-\frac12)}_{\lambda,0} e_5 +
\sqrt{\frac{i}{4\lambda}} \Psi^{(0,0)}_{\lambda+\frac12,-} e_1 + \sqrt{\frac{i}{4\lambda}}
\Psi^{(0,0)}_{\lambda-\frac12,+} e_1,\nonumber\\
&  G^{(6)}_{\lambda} = (i\lambda)^{1/2}\Psi^{(0,0)}_{\lambda,-} e_6 + \frac{1}{4\sqrt{\lambda}} \Big(i^{\frac32}(\lambda+\frac12)\Psi^{(\frac12,-\frac12)}_{\lambda+\frac12,-} (e_2+e_5)
+ i^{\frac12} \Psi^{(\frac12,\frac12)}_{\lambda+\frac12,-} (e_3-e_4)\Big) -\nonumber\\
& -\frac{1}{4\sqrt{\lambda}} \Big(i^{\frac32}(\lambda-\frac12)
\Psi^{(\frac12,-\frac12)}_{\lambda-\frac12,-} (-e_2-e_5) + i^{\frac12}
\Psi^{(\frac12,\frac12)}_{\lambda-\frac12,-}(e_3-e_4)\Big) + \frac{(i\lambda)^\frac{1}{2}}{\lambda^2-\frac14}
\Psi^{(0,0)}_{\lambda,-} e_1.\nonumber
\end{align}

The other set of solutions is similar and can be found in appendix A.
In this appendix we also demonstrate explicitly that the superconformal
blocks $G_\lambda$ can be mapped to the
superconformal blocks found in \cite{Cornagliotto:2017dup} by an appropriate gauge transformation.

\subsection{Shortening conditions}

In this final subsection we want to discuss shortening conditions and the relevant blocks for correlation
functions in which some of all of the external fields are BPS operators. Let us first consider the case of
the correlation function
\begin{equation}
\langle \varphi\mathcal{O}\bar\varphi\mathcal{O}\rangle, \label{eq:corr1}
\end{equation}
where the fields $\varphi$ and $\bar\varphi$ are chiral and anti-chiral, respectively, while the other two
fields $\mathcal{O}$ are arbitrary. This means that the operator $\varphi$ is annihilated by $Q_-$ and that
$\bar\varphi$ is annihilated by $Q_+$. Consequently, the superconformal blocks obey the following two shortening conditions
\begin{equation} \label{eq:constraintex1}
\mathcal{R}_{Q_-}G = \mathcal{L}_{Q_+}G = 0.
\end{equation}
As we have pointed out above right invariant vector fields for the fermionic generator $Q_-$ is simply
given by a partial derivative with respect to the coordinate $\bar\varrho$ while the left invariant vector
field for $Q_+$ is a derivative with respect to $\varrho$. In order for the constraints \eqref{eq:constraintex1}
to be satisfied, the conformal block should not have any dependence with respect to $\varrho,\bar\varrho$, i.e.\ it
should satisfy $G^{(i)}_j = 0$ for $j=3,4,5,6$. This is the case for two of our solutions, namely for the
blocks $G^{(1)}_n$ and $G^{(2)}_n$. This case further specialises to that of four short operators
\begin{equation}
\langle \varphi\varphi\bar\varphi\bar\varphi\rangle,  \label{eq:corr2}
\end{equation}
which is characterized by the additional two shortening conditions
\begin{equation}
\mathcal{R}_{s(Q_-)}G = \mathcal{L}_{s(Q_+)}G = 0. \label{eq:constraintex2}
\end{equation}
Since the Weyl inversion $s$ acts trivially on the $U(1)$ $R$-charge, the image of $Q_\pm$ is $S_\pm
= s(Q_\pm)$. Once again, the two vector fields that appear in eq.\ \eqref{eq:constraintex2} are given by
simple derivatives with respect to $\sigma$ and $\bar\sigma$. Relevant blocks should not possess any dependence on the
fermionic variables and hence $G^{(1)}$ is the only allowed solution in this case.

So far, all the shortening conditions we considered were of the simplest type in which right invariant
vector fields are taken with respect to elements in $\mathfrak{g}_-$ while left invariant vector fields
come with fermionic elements $X \in \mathfrak{g}_+$ of positive $R$-charge. Now we would like to consider
shortening for correlation functions of the form
\begin{equation}
\langle\varphi\bar\varphi\mathcal{O}\mathcal{O}\rangle. \label{eq:corr3}
\end{equation}
The case when the two BPS fields $\varphi$ and $\bar \varphi$ are in the third and fourth spot
is analogous. The differential equations obeyed by superconformal blocks now take
the form
\begin{equation}
\mathcal{R}_{Q_-}G = \mathcal{R}_{S_+}G = 0. \label{eq:constraintsex3}
\end{equation}
The first condition implies that $\bar\varrho$-components of the block vanish, as above. Looking back at our
expression for blocks $G^{(i)}$ we see that this condition is satisfied for $i = 1,2,3$. By pulling the
functions back to the supergroup $G$ we can explicitly check that the space of solutions of the second
condition is spanned by $\{G^{(2)},G^{(3)}\}$.

To verify that these two blocks do indeed satisfy $\mathcal{R}_{S_+} G=0$ is a bit cumbersome due to the
fact that right invariant vector fields with respect to fermionic directions of positive $R$-charge involve
differentiation with respect to both fermionic and bosonic coordinates with non-trivial coefficients. But the
answer we gave satisfies stringent consistency checks that arise from considering special cases of the
correlator \eqref{eq:corr3}. Let us first consider the case
\begin{equation}
\langle \varphi\bar\varphi\bar\varphi\varphi\rangle,\
\end{equation}
which is a specialization of the correlators \eqref{eq:corr1} and \eqref{eq:corr3} at the same time. Hence
the relevant block that satisfies
\begin{equation}
\mathcal{L}_{Q_+} G = \mathcal{L}_{S_-} G = 0,\
\end{equation}
in addition to the constraints \eqref{eq:constraintsex3}, so it must be $G^{(2)}$. The other special case addresses
correlators of the form
\begin{equation}
\langle \varphi\bar\varphi\varphi\bar \varphi\rangle,\
\end{equation}
which obey
\begin{equation}
\mathcal{L}_{Q_-} G = \mathcal{L}_{S_+} G = 0 .
\end{equation}
The second of equation tells us that the $\sigma$-components vanish and consequently the block
has to be $G^{(3)}$. Similarly, the blocks for correlation functions
\begin{equation}
\langle \bar\varphi\varphi\bar\varphi \varphi\rangle\ \text{and}\
\langle \bar\varphi\varphi\varphi\bar\varphi\rangle,
\end{equation}
are $G^{(4)}$ and $G^{(5)}$, respectively. This exhausts the possible shortening conditions.

\section{Summary, Outlook and Open Problems}

In this final section we want to provide a detailed outlook to some direct applications
of the Casimir equations we have derived above. In particular, we summarize the main
results of our upcoming paper on superblocks for $\mathcal{N}=1$ superconformal field
theories in $d=4$ dimensions. The section concludes with a list of directions that
should be pursued in the future.

\subsection{Brief summary of results}

Here we studied superconformal blocks, for both long and short external operators. In the first
step, we realized the space of blocks through a specific space of functions \eqref{eq:Gamma}
on the 2-dimensional bosonic double coset $K_\text{bos} \backslash G_\text{bos}/K_\text{bos}$
that take values in the finite dimensional space \eqref{eq:W} of four-point tensor structures.
In order to establish this model for the space of superblocks we proved a supersymmetric
extension of theorem 9.2 in \cite{Dobrev:1977qv}. Let us stress that this first step of our
analysis applies to all superconformal algebras, not just those of type I.

For superconformal algebras of type I we were able to obtain a universal formula
\eqref{eq:Casimir} for the Casimir operator. Its universality and the amazing simplicity is
based on a special choice \eqref{eq:variables} of coordinates that only exist for type I
superalgebras. The Casimir operator was written in the form of a quantum mechanical
Hamiltonian of Calogero-Sutherland type with a matrix valued potential. Unlike its
bosonic cousin, the potential of the supersymmetric model also involves hyperbolic
functions of the particle coordinates (cross ratios) that appear in the numerator.
The associated Schroedinger eigenvalue equations are equivalent to the Casimir
equations of superconformal field theory which had only been worked out in very
few cases before \cite{Cornagliotto:2017dup,Kos:2018glc}. Let us stress again that our Casimir operators for
type I superconformal algebras are completely explicit, at least for those cases
in which the Casimir equations of spinning blocks are known. The only term we
need in passing from spinning bosonic blocks to superblocks is the nilpotent
potential $A$ which contains matrix elements of a finite dimensional representation
of $SO(1,d+1)$, restricted to the 2-dimensional bosonic subspace $A$ in the $KAK$
decomposition of the bosonic conformal group $SO(1,d+1)$.

Because the additional term $A$ in the potential of the superconformal
Calogero-Sutherland model is nilpotent, one can construct eigenfunctions systematically
by means of a perturbative expansion that truncates at some finite order. We explained
this in general but only worked it out for one example, namely for $\mathcal{N}=2$
superconformal symmetry in $d=1$ dimensions. In this case we were able to recover
the known expressions for long superconformal blocks from \cite{Cornagliotto:2017dup}.
There are two points in this analysis that we would like to stress a bit more. The
first concerns our expressions for the integrals $I_1$ and $I_2$ in section 4.3 that
were needed to evaluate the perturbative expansion. One can rephrase these in terms
of representation theory of $SO(d+2) = SO(3)$ and express $I_1$ and $I_2$ through
certain Clebsch-Gordan coefficients. The generalization of this part of our analysis
will be developed in our forthcoming paper. It is clearly necessary in order to
obtain explicit formulas for superconformal blocks in dimension $d > 2$. Let us
stress, however, that this input into our program involves only results from the
group theory of compact groups $SO(d+2)$ and hence long established mathematics,
see e.g.\ \cite{Vilenkin}.

The second point in our example that was only explained for $\mathfrak{su}(1,1|1)$
and needs to be adapted for other superconformal symmetries is the map from wave
functions of the supersymmetric Calogero-Sutherland model to superconformal blocks.
In the example, this map was given in eqs.\ \eqref{eq:CStoCB1} and
\eqref{eq:CStoCB2}. We have pointed out before that this map does neither depend
on the weight and spin of the exchanged fields nor on the four-point tensor
structure. This feature makes it relatively easy to find the map for other
superconformal algebras on a case by case basis. Fundamentally, the map encodes
the isomorphism \eqref{eq:superMack} and can certainly be constructed
explicitly. We will address this in our forthcoming paper.

While we were able to write down the Casimir equations for superblocks explicitly,
even in cases without shortening, the size of the system of equations can become
very large. As we have seen above, shortening conditions can reduce the number $M$
of components and hence the complexity of the solution theory. Our approach allows
to implement at least part of the shortening conditions before solving the Casimir
equations. In order to obtain simple Casimir equations, we had to separate the
coordinates according to the $U(1)$ $R$-charge, placing those with positive $R$-charge
to the right and those with negative $R$-charge to the left. The left side is associated
with the fields that are inserted at $x_1$ and $x_2$. For these fields we can therefore
implement all shortening conditions that arise from supercharges with negative $U(1)$
$R$-charge. Shortening conditions for the fields at $x_3$ and $x_4$, on the other hand,
are easy to implement for supercharges with positive $R$-charge. Shortening conditions
involving supercharges with the opposite sign of the $R$-charge, however, can only be
imposed after solving the system. This means that, depending on the precise set of
external multiplets and on the channel, one may be forced to work with a set of
equations that is larger than the set of solutions.

\subsection{Applications to 4-dimensional $\mathcal{N}=1$ theories}

The physically most interesting cases to which our theory applies are $d=4$ dimensional superconformal
field theories. In a forthcoming paper we will work out explicit formulas for superconformal
blocks in theories with $\mathcal{N}=1$ supersymmetry. The blocks we shall consider appear
in the decomposition of four-point functions involving two long multiplets $\mathcal{O}$
along with one chiral field $\varphi$ and one anti-chiral $\bar \varphi$. Let us first
look at the four-point function
\begin{equation} \label{eq:d4correlation}
\langle \varphi(x_1) \mathcal{O}(x_2) \bar \varphi(x_3)\mathcal{O}(x_4) \rangle\ .
\end{equation}
The odd subspace $\mathfrak{g}_{(1)}$ of the Lie superalgebra $\mathfrak{g} = \mathfrak{sl}(4|1)$
has dimension $\text{dim}\mathfrak{g}_{(1)} = 8$ and hence the various subspaces $\mathfrak{q}_\pm$
and $\mathfrak{s}_\pm$ are all 2-dimensional. Under the action of $\mathfrak{k} = \mathfrak{so}
(1,1) \oplus \mathfrak{so}(4) \oplus \mathfrak{u}(1)$, these subspaces transform in the
representations
\begin{eqnarray}
\mathfrak{q}_+ & \cong & (1/2,(0,1/2),+) \ , \quad \mathfrak{s}_+ \cong (-1/2,(1/2,0),+)\ ,
\\[2mm]
\mathfrak{q}_- & \cong & (1/2,(1/2,0),-) \ , \quad \mathfrak{s}_- \cong (-1/2,(0,1/2),-)\ .
\end{eqnarray}
Here, the first label refers to the eigenvalue of dilations $D$, the pair in the second entry
describes a representation of $\mathfrak{so}(4) \cong \mathfrak{sl}(2) \oplus \mathfrak{sl}(2)$
and the final entry is the $U(1)$ R-charge. Upon restriction to the stabilizer subalgebra
$\mathfrak{b} \cong \mathfrak{so}(2) \oplus \mathfrak{u}(1)$, the 2-dimensional representations
split into two 1-dimensional ones in which the generator of the $\mathfrak{so}(2)$ assumes
opposite eigenvalues.

According to our general theory, blocks for the correlation function \eqref{eq:d4correlation}
are functions on the double coset with values in the space
\begin{equation}
\left( V_{(12)} \otimes \Lambda \mathfrak{s}_- \otimes V_{(34)} \otimes
\Lambda \mathfrak{s}_+ \right)^{\mathfrak{b}} = \left( \Lambda \mathfrak{s} \right)^\mathfrak{b}
\ . \label{eq:d4space}
\end{equation}
Here we assumed that all fields are scalar and that the $U(1)$ R-charge of the long multiplet
$\mathcal{O}$ vanishes. Under the action of $\mathfrak{k}$ the 16-dimensional exterior algebra
inside the brackets transforms as
\begin{eqnarray*}
	\Lambda \mathfrak{s} & \cong & (0,(0,0),0) \oplus (-1/2,(1/2,0),1) \oplus (-1/2,(0,1/2),-1) \\[2mm] & & \oplus (-1,(0,0),2) \oplus (-1,(0,0),-2)\oplus (-1,(1/2,1/2),0)
	\\[2mm] & & \oplus (-3/2,(1/2,0),-1) \oplus (-3/2,(0,1/2),1) \oplus (-2,(0,0),0)
\end{eqnarray*}
The first and the last term contain one $\mathfrak{b}$-invariant each and there are two more invariants in the last term of the second line - hence the dimension $M$
of the space \eqref{eq:d4space} is $M=4$. Consequently, the associated Casimir equation has four
components and it possesses four linearly independent solutions, in agreement with the claim in
a recent revision of \cite{Ramirez:2018lpd}. If $\mathcal{O}$ belongs to supermultiplet of a
conserved current, additional shortening conditions  reduce the number of components to $M$=3.
These shortening conditions can be implemented while setting up the equations so that the
solution theory simplifies further.

The Casimir equations for blocks of \eqref{eq:d4correlation} will be worked out and solved in our forthcoming
paper. It turns out that in this case the nilpotent term has order two as well, just as
in the case of the $d=1$ example we discussed in section 4. The eigenfunctions of the 
trigonometric version of the relevant Calogero-Sutherland model take the form
\begin{align*}
& f^{(1)}_l = \varphi^{(a,b)}_l e_0,\\[2mm] 
& f^{(2)}_l = \psi^{(a+\frac14,b-\frac14),i}_l e_i + \sum_{m\in l\otimes f} \Big(\frac{c_m}{E_l-E_m}\varphi^{(a,b)}_m \Big) e_0,\\[2mm]
& f^{(3)}_l = \varphi_l^{(a+\frac12,b-\frac12)} e_3 + \sum_{m \in l\otimes f,i}\frac{c^i_m}{E_l-E_m} \psi_m^{(a+\frac14,b-\frac14),i}e_i 
\\[2mm] & \hspace*{5cm} - \sum_{m,n}\Big(\frac{c_{mn}(E_m-E_n)}{(E_l-E_m)^2 (E_l-E_n)}\varphi^{(a,b)}_n\Big)e_0. 
\end{align*}
Here, $\varphi_l$ and $\psi^i_l,\ i=1,2$, are solutions to scalar and spin-1/2 models, derived in 
section 3.3 of \cite{Schomerus:2017eny}. There, it is also described how to map $\varphi$ and $\psi$ 
to scalar and seed conformal partial waves, respectively. Further, $l$ and $m$ label irreducible finite 
dimensional representations of the complexified conformal algebra $\mathfrak{so}(6)$, $f$ stands
for the fundamental representation of $\mathfrak{sl}(4)$, and the parameters $a,b$ are related to 
weights of the four fields by $2a = \Delta_2 - \Delta_1$ $2b=\Delta_3-\Delta_4$. The value of the 
quadratic Casimir in a representation $l$ is denoted by $E_l$, stressing its  role as energy in 
the associated quantum mechanics problem. Finally, the coefficients $c_m, c^i_m, c_{mn}$ are 
certain $\mathfrak{so}(6)$ Clebsch-Gordan coefficients which can be computed from the results 
existing in the literature. Explicit formulas for these coefficients, the extension to the 
hyperbolic model and the map to conformal blocks in 4-dimensional $\mathcal{N}=1$ theories 
will be described in our forthcoming publication where we shall also compare our results with 
the recent formulas in \cite{Ramirez:2018lpd}.  

Blocks for the second channel in which we form the pairs $(\varphi \bar \varphi)$ and $(\mathcal{O}
\mathcal{O})$ are a little more difficult to obtain since we can only impose half of the shortening
conditions before working out and solving the Casimir equation. The second half of the shortening
conditions is then imposed on the solutions, following the same logic we outlined in section 4.5.
The original Casimir equation acts on functions that take values in the space
$$ \left( \Lambda (\mathfrak{q_+} \oplus \mathfrak{s}) \right)^\mathfrak{b} \ . $$
Using the input we spelled out above, it is straightforward to compute that this space has
dimension $M = 9$. Hence our original system of Casimir equations has nine solutions, five
of which are eliminated when we impose the remaining two shortening conditions. Once again,
it is possible to construct these solutions explicitly.

\subsection{Some further directions}

There are many further directions that would be interesting to study in the future. In the case
of $4$-dimensional $\mathcal{N}=1$ theories, one of the most interesting correlators is the one
of four conserved currents $J$. The corresponding multiplet satisfies some shortening conditions,
but is not half-BPS. In this case we can implement half of the shortening conditions before we
set up the Casimir equations. It turns out that this step reduces the number of components of
the Casimir equations from $M=36$, the number one needs to study generic long multiplets of vanishing total $R$-charge, down
to $M=15$. This is still a fairly large system, but we expect that the nilpotent perturbation
theory is manageable. Once the solutions have been constructed, the second half of the
shortening conditions are used so select the relevant subset from the $15$ solutions one finds
while solving the partially shortened Casimir system.

Given the enormous interest in $4$-dimensional $\mathcal{N}=2$ superconformal field theories,
it would certainly be desirable to construct superblocks for $\mathfrak{su}(2,2|2)$. Once again,
it is a good strategy to consider e.g. mixed correlators between BPS operators and longer
multiplets, and in particular of correlators involving the stress tensor multiplet. In
$\mathcal{N}=2$ theories, there exist two types of half-BPS fields. Coulomb branch operators
$\mathcal{E}_r$ satisfy two shortening conditions which have the same sign of $R$-charge and
hence there is a chance to implement their shortening before solving the Casimir equations.
Higgs Branch operators $\hat{\mathcal{B}}_R$, on the other hand, are characterized by two
shortening conditions of opposite $U(1)$ $R$-charge and  hence only half of the shortening
conditions may be implemented before we solve the Casimir equations.

Let us stress again that even in the absence of shortening, our solution theory for the Casimir 
equations is algorithmic and gives results in terms of finite sums over spinning bosonic blocks. 
The only issue we face if we can only implement a small number of shortening conditions before 
solving the equations is that the solutions involve a larger sum of terms. These sums may obscure 
some features of the new superblocks. One way to understand the properties of blocks is through 
the integrability of the  Calogero-Sutherland model \cite{Isachenkov:2017qgn}. Integrability of the 
supersymmetric Casimir equations for superconformal algebras therefore offers another interesting 
direction for future research. Ordinary Calogero-Sutherland systems are well known to be integrable 
and even super-integrable in some cases, see \cite{Isachenkov:2017qgn,Schomerus:2017eny}
for an extensive list of references. Certain types of models with spin have recently been studied 
in \cite{Reshetikhin:2015rba,Reshetikhin} from the Hamiltonian reduction perspective and their 
degenerate (or super-) integrability was established. The quantum analogue of \cite{Reshetikhin} 
finds roots in the work of Harish-Chandra and Kostant and Tirao \cite{Kostant}. One observes that 
our reduction of section 2 is a supersymmetric version of these constructions. The algebra of first 
integrals is generally enlarged compared to the bosonic one - already in the model that was solved 
in section 4.3 it contains the supersymmetry algebra $\mathfrak{sl}(1|1)$. It seems that similar 
statements hold for other superconformal algebras. We will return to this topic in future work.

Finally, it would also be interesting to study two-point functions of local superfields in the 
presence of a superconformal defect. Such a configuration admits two channels, the defect and 
the bulk channel \cite{Billo:2016cpy,Lauria:2017wav,Isachenkov:2018pef,Lauria:2018klo}. While 
defect channel blocks split into simple products of hypergeometric functions in one variable
\cite{Billo:2016cpy}, the construction for bulk channel blocks with scalar external fields 
was only completed recently in \cite{Isachenkov:2018pef}, see also \cite{Lauria:2017wav} for 
earlier results. An extension to bulk channel blocks for spinning operators was initiated 
recently in \cite{Lauria:2018klo}. The extension of such studies to superconformal defects 
should be simpler than for four-point functions of bulk fields. On the one hand, the smaller 
number of tensor structures in the defect setup reduces the size of the system of Casimir 
equations as well as the order to which we have to perform our perturbative expansion. In 
addition, the starting point of our perturbation series, spinning bosonic defect blocks, 
are also simpler than their four-point counterparts \cite{Lauria:2018klo}. There are 
many potentially interesting applications, in particular to line defects in superconformal 
field theories, see e.g.\ \cite{Balitsky:2013npa,Balitsky:2015tca,Balitsky:2015oux,Liendo:2018ukf}.  
\bigskip

\noindent
{\bf Acknowledgement:} We thank Martina Cornagliotto, James Drummond, Aleix Gimenez, Misha Isa\-chen\-kov, Sylvain
Lacroix and in particular Madalena Lemos, Pedro Liendo and Maja Buri\' c for comments and very
fruitful discussion. The work of ES was supported by ERC grant 648630 IQFT and in part by the grant "Exact Results in Gauge and String Theories" from the Knut and Alice Wallenberg foundation.

\appendix

\section{Conformal partial waves and blocks for $\mathfrak{sl}(2|1)$}

In this appendix, we collect some formulas which complement the derivation of superconformal blocks from compact partial waves given in section 4.4. The solutions obtained by analytically continuing $f_n$ to complex values of the parameter read

\begin{align*}
& F^{(1)}_{\lambda} = \Psi_\lambda e_1,\\
& F^{(2)}_{\lambda} = \Phi_\lambda e_2 - \frac{1}{\sqrt{4\lambda(\lambda+\frac12)}}
\Psi_{\lambda+\frac12} e_1 - \frac{1}{\sqrt{4\lambda (\lambda - \frac12)}} \Psi_{\lambda-\frac12} e_1,\\
& F^{(3)}_{\lambda} = X_\lambda e_3 + \frac{1}{\sqrt{4\lambda(\lambda+\frac12)}}
\Psi_{\lambda+\frac12} e_1 - \frac{1}{\sqrt{4\lambda (\lambda - \frac12)}} \Psi_{\lambda-\frac12} e_1,\\
& F^{(4)}_{\lambda} = X_\lambda e_4 - \frac{1}{\sqrt{4\lambda(\lambda+\frac12)}}
\Psi_{\lambda+\frac12} e_1 + \frac{1}{\sqrt{4\lambda (\lambda - \frac12)}} \Psi_{\lambda-\frac12} e_1,\\
& F^{(5)}_{\lambda} = \Phi_\lambda e_5 - \frac{1}{\sqrt{4\lambda(\lambda+\frac12)}}
\Psi_{\lambda+\frac12} e_1 - \frac{1}{\sqrt{4\lambda (\lambda - \frac12)}} \Psi_{\lambda-\frac12} e_1,\\
& F^{(6)}_{\lambda} = \Psi_\lambda e_6 + \frac{1}{\sqrt{4\lambda(\lambda+\frac12)}}
\Big(\Phi_{\lambda+\frac12} (-e_2-e_5) + X_{\lambda+\frac12} (-e_3+e_4)\Big) -\\
& -\frac{1}{\sqrt{4\lambda (\lambda - \frac12)}} \Big(\Phi_{\lambda-\frac12} (e_2+e_5) +
X_{\lambda-\frac12}(-e_3+e_4)\Big) + \frac{1}{\lambda^2-\frac14} \Psi_\lambda e_1.
\end{align*}
Notice that $F^{(6)}_{\lambda}$ comes from the analytic continuation of a linear combination of $f^{(1)}_n$ and $f^{(6)}_n$.
Our choice ensures that functions $F^{(i)}_{\lambda}$ are eigenfunctions of both quadratic and cubic Casimirs.

A family of superconformal blocks is obtained from these wavefunctions by reexpressing all the functions in terms of $\Psi^{(a,b)}_\lambda$ with the help of eqs.\ \eqref{eq:functions} and then attach a subscript $+$
\begin{align*}
&  G^{(1)}_{+,\lambda} = (i\lambda)^{1/2} \Psi^{(0,0)}_{\lambda,+} e_1,\\
&  G^{(2)}_{+,\lambda} = \frac12 (i\lambda)^{3/2} \Psi^{(\frac12,-\frac12)}_{\lambda,+} e_2 -
\sqrt{\frac{i}{4\lambda}} \Psi^{(0,0)}_{\lambda+\frac12,+} e_1 - \sqrt{\frac{i}{4\lambda}}
\Psi^{(0,0)}_{\lambda-\frac12,+} e_1 ,\\
&  G^{(3)}_{+,\lambda} = \frac12 (i\lambda)^{1/2} \Psi^{(\frac12,\frac12)}_{\lambda,+} e_3 +
\sqrt{\frac{i}{4\lambda}} \Psi^{(0,0)}_{\lambda+\frac12,+} e_1 - \sqrt{\frac{i}{4\lambda}}
\Psi^{(0,0)}_{\lambda-\frac12,+} e_1,\\
&  G^{(4)}_{+,\lambda} = \frac12 (i\lambda)^{1/2} \Psi^{(\frac12,\frac12)}_{\lambda,+} e_4 -
\sqrt{\frac{i}{4\lambda}} \Psi^{(0,0)}_{\lambda+\frac12,+} e_1 + \sqrt{\frac{i}{4\lambda}}
\Psi^{(0,0)}_{\lambda-\frac12,+} e_1,\\
&  G^{(5)}_{+,\lambda} = \frac12 (i\lambda)^{3/2} \Psi^{(\frac12,-\frac12)}_{\lambda,0} e_5 -
\sqrt{\frac{i}{4\lambda}} \Psi^{(0,0)}_{\lambda+\frac12,+} e_1 - \sqrt{\frac{i}{4\lambda}}
\Psi^{(0,0)}_{\lambda-\frac12,+} e_1,\\
&  G^{(6)}_{+,\lambda} = (i\lambda)^{1/2}\Psi^{(0,0)}_{\lambda,+} e_6 + \frac{\sqrt{\lambda}}{4\lambda} \Big(i^{\frac32}(\lambda+\frac12)\Phi^{(\frac12,-\frac12)}_{\lambda+\frac12,+} (-e_2-e_5)
+ i^{\frac12} \Psi^{(\frac12,\frac12)}_{\lambda+\frac12,+} (-e_3+e_4)\Big) -\\
& -\frac{\sqrt{\lambda}}{4\lambda}\Big(i^{\frac32}(\lambda-\frac12)
\Psi^{(\frac12,-\frac12)}_{\lambda-\frac12,+} (e_2+e_5) + i^{\frac12}
\Psi^{(\frac12,\frac12)}_{\lambda-\frac12,+}(-e_3+e_4)\Big) + \frac{(i\lambda)^{1/2} }{\lambda^2-\frac14}
\Psi^{(0,0)}_{\lambda,+} e_1.
\end{align*}
These solutions are valid for $\lambda<0$ and are related to the solutions $(\ref{solutions})$ by
\begin{equation}
G^{(j)}_{+,-\lambda}=i G^{(j)}_{-,\lambda},\ j\in\{1,3,4,6\},\ \ \ G^{(k)}_{+,-\lambda}=-i G^{(k)}_{-,\lambda},\ k\in\{2,5\}.
\end{equation}
Finally, we will map this system of solutions of the superconformal Casimir equations to the blocks that were constructed in \cite{Cornagliotto:2017dup}. To ease notation, we will drop the subscripts $+,\lambda$. The transformation between the two systems of functions is independent of the choice of tensor structures and of the exchanged field, i.e.\ it does neither depend on the parameter $\lambda$ nor on the superscript $(i)$ that labels the six independent solutions. We describe this transformation in two steps. The first one
can be described as a matrix valued gauge transformation of the form
\begin{align}
& G_1 = -4\omega g_0, \nonumber \\[2mm]
& G_2 = 2\omega \sqrt{z}\big(g_1 + g_2 - (z-1)g_4),\nonumber \\[2mm]
& G_3 = 2\omega  \sqrt{z(z-1)} \big(g_1-g_2+\frac{1}{z-1}g_3\big),\label{eq:CStoCB1}\\[2mm]
& G_4 = -2\omega \sqrt{z(z-1)} \big(g_1+g_2+g_4\big),\nonumber \\[2mm]
& G_5 = 2\omega\sqrt{z} \big(g_1-g_2-g_3\big),\nonumber \\[2mm]
& G_6 =  \omega z^2g_5, \nonumber
\end{align}
where $z = - \sinh^{-2} (u/2) = y/(y-1)$ and $\omega^2=\sin{2\mu}$ is the factor coming form the Haar measure. For each of the six solutions $G^{(i)} = (G^{(i)}_j,j=1, \dots,6)$ with
$i = 1, \dots, 6$ of our matrix valued Calogero-Sutherland model, the above transformation allows us to construct
a new set of functions $g_e = g^{(i)}_e, e = 0, \dots, 5$. The second step now amounts to a redefinition of the
bosonic conformal invariants. Given the set $g_e$ we claim that the following functions
\begin{align}
& f_0=g_0,\ \ f_1=g_1,\ \ f_2=g_2, \nonumber \\[2mm]
& f_3=g_3+(z-1)g_0',\ \ f_4=g_4+g_0',\label{eq:CStoCB2}\\[2mm]
& f_5=g_5 + g_3' + g_4 + (z-1) g_4' + g_0' + (z-1)g_0'', \nonumber
\end{align}
coincide with the functions $f_e, e=0, \dots, 5$ that were introduced in \cite{Cornagliotto:2017dup}. If we apply these transformations to our first solutions $G^{(1)}$, for example, the resulting set of $f_e$ are those blocks that describe the intermediate channel in which the $U(1)$ charge is shifted by one unit. Let us note that the transformation we have constructed is invertible and can also be used in order to construct our blocks $G_i$ from the blocks $f_e$ in \cite{Cornagliotto:2017dup}. The relation between the eigenvalues is $\lambda=h_{ex}$ and $c_3/\varepsilon=q_{ex}$.


\begin{thebibliography}{10}
	\providecommand{\href}[2]{#2}
	\providecommand{\arxivref}[2]{\href{http://arxiv.org/abs/#1}{#2}}
	\providecommand{\doiref}[2]{\href{http://dx.doi.org/#1}{#2}}
	\providecommand{\nbbstauthor}[1]{#1}
	\providecommand{\nbbstjournal}[1]{\textsf{#1}}
	\providecommand{\nbbsttitle}[1]{\textit{``#1''}}
	\providecommand{\nbbsturl}[1]{\texttt{#1}}
	\providecommand{\nbbsteprint}[1]{\texttt{#1}}
	\providecommand{\nbbststyle}{\raggedright\small\parskip0pt}
	\nbbststyle
	
	\bibitem{Dolan:2000ut}
	\nbbstauthor{F.~Dolan and H.~Osborn},
	\nbbsttitle{{Conformal four point functions and the operator product
			expansion}},
	\nbbstjournal{\doiref{10.1016/S0550-3213(01)00013-X}{Nucl.Phys.~B599,~459~(2001)}},
	\nbbsteprint{\arxivref{hep-th/0011040}{hep-th/0011040}}.
	%%CITATION = HEP-TH/0011040;%%
	
	\bibitem{Dolan:2003hv}
	\nbbstauthor{F.~Dolan and H.~Osborn},
	\nbbsttitle{{Conformal partial waves and the operator product expansion}},
	\nbbstjournal{\doiref{10.1016/j.nuclphysb.2003.11.016}{Nucl.Phys.~B678,~491~(2004)}},
	\nbbsteprint{\arxivref{hep-th/0309180}{hep-th/0309180}}.
	%%CITATION = HEP-TH/0309180;%%
	
	\bibitem{Ferrara:1972uq}
	\nbbstauthor{S.~Ferrara, A.~F.~Grillo, G.~Parisi and R.~Gatto},
	\nbbsttitle{{The shadow operator formalism for conformal algebra. Vacuum
			expectation values and operator products}},
	\nbbstjournal{\doiref{10.1007/BF02907130}{Lett.~Nuovo~Cim.~4S2,~115~(1972)}},
	[Lett. Nuovo Cim.4,115(1972)].
	%%CITATION = NCLTA,4S2,115;%%
	
	\bibitem{Rattazzi:2008pe}
	\nbbstauthor{R.~Rattazzi, V.~S.~Rychkov, E.~Tonni and A.~Vichi},
	\nbbsttitle{{Bounding scalar operator dimensions in 4D CFT}},
	\nbbstjournal{\doiref{10.1088/1126-6708/2008/12/031}{JHEP~0812,~031~(2008)}},
	\nbbsteprint{\arxivref{0807.0004}{arxiv:0807.0004}}.
	%%CITATION = ARXIV:0807.0004;%%
	
	\bibitem{ElShowk:2012ht}
	\nbbstauthor{S.~El-Showk, M.~F.~Paulos, D.~Poland, S.~Rychkov,
		D.~Simmons-Duffin et~al.},
	\nbbsttitle{{Solving the 3D Ising Model with the Conformal Bootstrap}},
	\nbbstjournal{\doiref{10.1103/PhysRevD.86.025022}{Phys.Rev.~D86,~025022~(2012)}},
	\nbbsteprint{\arxivref{1203.6064}{arxiv:1203.6064}}.
	%%CITATION = ARXIV:1203.6064;%%
	
	\bibitem{El-Showk:2014dwa}
	\nbbstauthor{S.~El-Showk, M.~F.~Paulos, D.~Poland, S.~Rychkov,
		D.~Simmons-Duffin et~al.},
	\nbbsttitle{{Solving the 3d Ising Model with the Conformal Bootstrap II.
			c-Minimization and Precise Critical Exponents}},
	\nbbstjournal{\doiref{10.1007/s10955-014-1042-7}{J.Stat.Phys.~xx,~xx~(2014)}},
	\nbbsteprint{\arxivref{1403.4545}{arxiv:1403.4545}}.
	%%CITATION = ARXIV:1403.4545;%%
	
	\bibitem{Simmons-Duffin:2015qma}
	\nbbstauthor{D.~Simmons-Duffin},
	\nbbsttitle{{A Semidefinite Program Solver for the Conformal Bootstrap}},
	\nbbstjournal{\doiref{10.1007/JHEP06(2015)174}{JHEP~1506,~174~(2015)}},
	\nbbsteprint{\arxivref{1502.02033}{arxiv:1502.02033}}.
	%%CITATION = ARXIV:1502.02033;%%
	
	\bibitem{Kos:2016ysd}
	\nbbstauthor{F.~Kos, D.~Poland, D.~Simmons-Duffin and A.~Vichi},
	\nbbsttitle{{Precision Islands in the Ising and $O(N)$ Models}},
	\nbbsteprint{\arxivref{1603.04436}{arxiv:1603.04436}}.
	%%CITATION = ARXIV:1603.04436;%%
	
	\bibitem{Dolan:2011dv}
	\nbbstauthor{F.~Dolan and H.~Osborn},
	\nbbsttitle{{Conformal Partial Waves: Further Mathematical Results}},
	\nbbsteprint{\arxivref{1108.6194}{arxiv:1108.6194}}.
	%%CITATION = ARXIV:1108.6194;%%
	
	\bibitem{Costa:2011dw}
	\nbbstauthor{M.~S.~Costa, J.~Penedones, D.~Poland and S.~Rychkov},
	\nbbsttitle{{Spinning Conformal Blocks}},
	\nbbstjournal{\doiref{10.1007/JHEP11(2011)154}{JHEP~1111,~154~(2011)}},
	\nbbsteprint{\arxivref{1109.6321}{arxiv:1109.6321}}.
	%%CITATION = ARXIV:1109.6321;%%
	
	\bibitem{SimmonsDuffin:2012uy}
	\nbbstauthor{D.~Simmons-Duffin},
	\nbbsttitle{{Projectors, Shadows, and Conformal Blocks}},
	\nbbstjournal{\doiref{10.1007/JHEP04(2014)146}{JHEP~1404,~146~(2014)}},
	\nbbsteprint{\arxivref{1204.3894}{arxiv:1204.3894}}.
	%%CITATION = ARXIV:1204.3894;%%
	
	\bibitem{Hogervorst:2013sma}
	\nbbstauthor{M.~Hogervorst and S.~Rychkov},
	\nbbsttitle{{Radial Coordinates for Conformal Blocks}},
	\nbbstjournal{\doiref{10.1103/PhysRevD.87.106004}{Phys.~Rev.~D87,~106004~(2013)}},
	\nbbsteprint{\arxivref{1303.1111}{arxiv:1303.1111}}.
	%%CITATION = ARXIV:1303.1111;%%
	
	\bibitem{Penedones:2015aga}
	\nbbstauthor{J.~Penedones, E.~Trevisani and M.~Yamazaki},
	\nbbsttitle{{Recursion Relations for Conformal Blocks}},
	\nbbsteprint{\arxivref{1509.00428}{arxiv:1509.00428}}.
	%%CITATION = ARXIV:1509.00428;%%
	
	\bibitem{Costa:2016xah}
	\nbbstauthor{M.~Costa, T.~Hansen, J.~Penedones and E.~Trevisani},
	\nbbsttitle{{Radial expansion for spinning conformal blocks}},
	\nbbsteprint{\arxivref{1603.05552}{arxiv:1603.05552}}.
	%%CITATION = ARXIV:1603.05552;%%
	
	\bibitem{Echeverri:2016dun}
	\nbbstauthor{A.~Castedo~Echeverri, E.~Elkhidir, D.~Karateev and M.~Serone},
	\nbbsttitle{{Seed Conformal Blocks in 4D CFT}},
	\nbbstjournal{\doiref{10.1007/JHEP02(2016)183}{JHEP~1602,~183~(2016)}},
	\nbbsteprint{\arxivref{1601.05325}{arxiv:1601.05325}}.
	%%CITATION = ARXIV:1601.05325;%%
	
	\bibitem{Cuomo:2017wme}
	\nbbstauthor{G.~F.~Cuomo, D.~Karateev and P.~Kravchuk},
	\nbbsttitle{{General Bootstrap Equations in 4D CFTs}},
	\nbbstjournal{\doiref{10.1007/JHEP01(2018)130}{JHEP~1801,~130~(2018)}},
	\nbbsteprint{\arxivref{1705.05401}{arxiv:1705.05401}},
	[,57(2017)].
	%%CITATION = ARXIV:1705.05401;%%
	
	\bibitem{Isachenkov:2016gim}
	\nbbstauthor{M.~Isachenkov and V.~Schomerus},
	\nbbsttitle{{Superintegrability of $d$-dimensional Conformal Blocks}},
	\nbbstjournal{\doiref{10.1103/PhysRevLett.117.071602}{Phys.~Rev.~Lett.~117,~071602~(2016)}},
	\nbbsteprint{\arxivref{1602.01858}{arxiv:1602.01858}}.
	%%CITATION = ARXIV:1602.01858;%%
	
	\bibitem{Heckman-Opdam}
	\nbbstauthor{G.~J.~Heckman and E.~M.~Opdam},
	\nbbsttitle{Root systems and hypergeometric functions. I},
	\nbbstjournal{Compositio~Mathematica~64,~329~(1987)}.
	
	\bibitem{Schomerus:2016epl}
	\nbbstauthor{V.~Schomerus, E.~Sobko and M.~Isachenkov},
	\nbbsttitle{{Harmony of Spinning Conformal Blocks}},
	\nbbstjournal{\doiref{10.1007/JHEP03(2017)085}{JHEP~1703,~085~(2017)}},
	\nbbsteprint{\arxivref{1612.02479}{arxiv:1612.02479}}.
	%%CITATION = ARXIV:1612.02479;%%
	
	\bibitem{Schomerus:2017eny}
	\nbbstauthor{V.~Schomerus and E.~Sobko},
	\nbbsttitle{{From Spinning Conformal Blocks to Matrix Calogero-Sutherland
			Models}},
	\nbbstjournal{\doiref{10.1007/JHEP04(2018)052}{JHEP~1804,~052~(2018)}},
	\nbbsteprint{\arxivref{1711.02022}{arxiv:1711.02022}}.
	%%CITATION = ARXIV:1711.02022;%%
	
	\bibitem{Isachenkov:2018pef}
	\nbbstauthor{M.~Isachenkov, P.~Liendo, Y.~Linke and V.~Schomerus},
	\nbbsttitle{{Calogero-Sutherland Approach to Defect Blocks}},
	\nbbstjournal{\doiref{10.1007/JHEP10(2018)204}{JHEP~1810,~204~(2018)}},
	\nbbsteprint{\arxivref{1806.09703}{arxiv:1806.09703}}.
	%%CITATION = ARXIV:1806.09703;%%
	
	\bibitem{Beem:2013qxa}
	\nbbstauthor{C.~Beem, L.~Rastelli and B.~C.~van~Rees},
	\nbbsttitle{{The $\mathcal N=4$ Superconformal Bootstrap}},
	\nbbstjournal{\doiref{10.1103/PhysRevLett.111.071601}{Phys.~Rev.~Lett.~111,~071601~(2013)}},
	\nbbsteprint{\arxivref{1304.1803}{arxiv:1304.1803}}.
	%%CITATION = ARXIV:1304.1803;%%
	
	\bibitem{Beem:2014zpa}
	\nbbstauthor{C.~Beem, M.~Lemos, P.~Liendo, L.~Rastelli and B.~C.~van~Rees},
	\nbbsttitle{{The $ \mathcal{N}=2 $ superconformal bootstrap}},
	\nbbstjournal{\doiref{10.1007/JHEP03(2016)183}{JHEP~1603,~183~(2016)}},
	\nbbsteprint{\arxivref{1412.7541}{arxiv:1412.7541}}.
	%%CITATION = ARXIV:1412.7541;%%
	
	\bibitem{Beem:2016wfs}
	\nbbstauthor{C.~Beem, L.~Rastelli and B.~C.~van~Rees},
	\nbbsttitle{{More ${\mathcal N}=4$ superconformal bootstrap}},
	\nbbstjournal{\doiref{10.1103/PhysRevD.96.046014}{Phys.~Rev.~D96,~046014~(2017)}},
	\nbbsteprint{\arxivref{1612.02363}{arxiv:1612.02363}}.
	%%CITATION = ARXIV:1612.02363;%%
	
	\bibitem{Dolan:2001tt}
	\nbbstauthor{F.~A.~Dolan and H.~Osborn},
	\nbbsttitle{{Superconformal symmetry, correlation functions and the operator
			product expansion}},
	\nbbstjournal{\doiref{10.1016/S0550-3213(02)00096-2}{Nucl.~Phys.~B629,~3~(2002)}},
	\nbbsteprint{\arxivref{hep-th/0112251}{hep-th/0112251}}.
	%%CITATION = HEP-TH/0112251;%%
	
	\bibitem{Dolan:2004mu}
	\nbbstauthor{F.~A.~Dolan, L.~Gallot and E.~Sokatchev},
	\nbbsttitle{{On four-point functions of 1/2-BPS operators in general
			dimensions}},
	\nbbstjournal{\doiref{10.1088/1126-6708/2004/09/056}{JHEP~0409,~056~(2004)}},
	\nbbsteprint{\arxivref{hep-th/0405180}{hep-th/0405180}}.
	%%CITATION = HEP-TH/0405180;%%
	
	\bibitem{Nirschl:2004pa}
	\nbbstauthor{M.~Nirschl and H.~Osborn},
	\nbbsttitle{{Superconformal Ward identities and their solution}},
	\nbbstjournal{\doiref{10.1016/j.nuclphysb.2005.01.013}{Nucl.~Phys.~B711,~409~(2005)}},
	\nbbsteprint{\arxivref{hep-th/0407060}{hep-th/0407060}}.
	%%CITATION = HEP-TH/0407060;%%
	
	\bibitem{Fortin:2011nq}
	\nbbstauthor{J.-F.~Fortin, K.~Intriligator and A.~Stergiou},
	\nbbsttitle{{Current OPEs in Superconformal Theories}},
	\nbbstjournal{\doiref{10.1007/JHEP09(2011)071}{JHEP~1109,~071~(2011)}},
	\nbbsteprint{\arxivref{1107.1721}{arxiv:1107.1721}}.
	%%CITATION = ARXIV:1107.1721;%%
	
	\bibitem{Fitzpatrick:2014oza}
	\nbbstauthor{A.~L.~Fitzpatrick, J.~Kaplan, Z.~U.~Khandker, D.~Li, D.~Poland and
		D.~Simmons-Duffin},
	\nbbsttitle{{Covariant Approaches to Superconformal Blocks}},
	\nbbstjournal{\doiref{10.1007/JHEP08(2014)129}{JHEP~1408,~129~(2014)}},
	\nbbsteprint{\arxivref{1402.1167}{arxiv:1402.1167}}.
	%%CITATION = ARXIV:1402.1167;%%
	
	\bibitem{Bissi:2015qoa}
	\nbbstauthor{A.~Bissi and T.~Łukowski},
	\nbbsttitle{{Revisiting $ \mathcal{N}=4 $ superconformal blocks}},
	\nbbstjournal{\doiref{10.1007/JHEP02(2016)115}{JHEP~1602,~115~(2016)}},
	\nbbsteprint{\arxivref{1508.02391}{arxiv:1508.02391}}.
	%%CITATION = ARXIV:1508.02391;%%
	
	\bibitem{Lemos:2015awa}
	\nbbstauthor{M.~Lemos and P.~Liendo},
	\nbbsttitle{{Bootstrapping $ \mathcal{N}=2 $ chiral correlators}},
	\nbbstjournal{\doiref{10.1007/JHEP01(2016)025}{JHEP~1601,~025~(2016)}},
	\nbbsteprint{\arxivref{1510.03866}{arxiv:1510.03866}}.
	%%CITATION = ARXIV:1510.03866;%%
	
	\bibitem{Liendo:2016ymz}
	\nbbstauthor{P.~Liendo and C.~Meneghelli},
	\nbbsttitle{{Bootstrap equations for $ \mathcal{N} $ = 4 SYM with defects}},
	\nbbstjournal{\doiref{10.1007/JHEP01(2017)122}{JHEP~1701,~122~(2017)}},
	\nbbsteprint{\arxivref{1608.05126}{arxiv:1608.05126}}.
	%%CITATION = ARXIV:1608.05126;%%
	
	\bibitem{Lemos:2016xke}
	\nbbstauthor{M.~Lemos, P.~Liendo, C.~Meneghelli and V.~Mitev},
	\nbbsttitle{{Bootstrapping $\mathcal{N}=3$ superconformal theories}},
	\nbbstjournal{\doiref{10.1007/JHEP04(2017)032}{JHEP~1704,~032~(2017)}},
	\nbbsteprint{\arxivref{1612.01536}{arxiv:1612.01536}}.
	%%CITATION = ARXIV:1612.01536;%%
	
	\bibitem{Chang:2017xmr}
	\nbbstauthor{C.-M.~Chang and Y.-H.~Lin},
	\nbbsttitle{{Carving Out the End of the World or (Superconformal Bootstrap in
			Six Dimensions)}},
	\nbbstjournal{\doiref{10.1007/JHEP08(2017)128}{JHEP~1708,~128~(2017)}},
	\nbbsteprint{\arxivref{1705.05392}{arxiv:1705.05392}}.
	%%CITATION = ARXIV:1705.05392;%%
	
	\bibitem{Liendo:2018ukf}
	\nbbstauthor{P.~Liendo, C.~Meneghelli and V.~Mitev},
	\nbbsttitle{{Bootstrapping the half-BPS line defect}},
	\nbbstjournal{\doiref{10.1007/JHEP10(2018)077}{JHEP~1810,~077~(2018)}},
	\nbbsteprint{\arxivref{1806.01862}{arxiv:1806.01862}}.
	%%CITATION = ARXIV:1806.01862;%%
	
	\bibitem{Khandker:2014mpa}
	\nbbstauthor{Z.~U.~Khandker, D.~Li, D.~Poland and D.~Simmons-Duffin},
	\nbbsttitle{{$ \mathcal{N} $ = 1 superconformal blocks for general scalar
			operators}},
	\nbbstjournal{\doiref{10.1007/JHEP08(2014)049}{JHEP~1408,~049~(2014)}},
	\nbbsteprint{\arxivref{1404.5300}{arxiv:1404.5300}}.
	%%CITATION = ARXIV:1404.5300;%%
	
	\bibitem{Berkooz:2014yda}
	\nbbstauthor{M.~Berkooz, R.~Yacoby and A.~Zait},
	\nbbsttitle{{Bounds on $\mathcal{N} = 1$ superconformal theories with global
			symmetries}},
	\nbbstjournal{\doiref{10.1007/JHEP01(2015)132,
			10.1007/JHEP08(2014)008}{JHEP~1408,~008~(2014)}},
	\nbbsteprint{\arxivref{1402.6068}{arxiv:1402.6068}},
	[Erratum: JHEP01,132(2015)].
	%%CITATION = ARXIV:1402.6068;%%
	
	\bibitem{Li:2016chh}
	\nbbstauthor{Z.~Li and N.~Su},
	\nbbsttitle{{The Most General $4\mathcal{D}$ $\mathcal{N}=1$ Superconformal
			Blocks for Scalar Operators}},
	\nbbstjournal{\doiref{10.1007/JHEP05(2016)163}{JHEP~1605,~163~(2016)}},
	\nbbsteprint{\arxivref{1602.07097}{arxiv:1602.07097}}.
	%%CITATION = ARXIV:1602.07097;%%
	
	\bibitem{Li:2017ddj}
	\nbbstauthor{D.~Li, D.~Meltzer and A.~Stergiou},
	\nbbsttitle{{Bootstrapping mixed correlators in 4D $ \mathcal{N} $ = 1 SCFTs}},
	\nbbstjournal{\doiref{10.1007/JHEP07(2017)029}{JHEP~1707,~029~(2017)}},
	\nbbsteprint{\arxivref{1702.00404}{arxiv:1702.00404}}.
	%%CITATION = ARXIV:1702.00404;%%
	
	\bibitem{Doobary:2015gia}
	\nbbstauthor{R.~Doobary and P.~Heslop},
	\nbbsttitle{{Superconformal partial waves in Grassmannian field theories}},
	\nbbstjournal{\doiref{10.1007/JHEP12(2015)159}{JHEP~1512,~159~(2015)}},
	\nbbsteprint{\arxivref{1508.03611}{arxiv:1508.03611}}.
	%%CITATION = ARXIV:1508.03611;%%
	
	\bibitem{Aprile:2017bgs}
	\nbbstauthor{F.~Aprile, J.~M.~Drummond, P.~Heslop and H.~Paul},
	\nbbsttitle{{Quantum Gravity from Conformal Field Theory}},
	\nbbstjournal{\doiref{10.1007/JHEP01(2018)035}{JHEP~1801,~035~(2018)}},
	\nbbsteprint{\arxivref{1706.02822}{arxiv:1706.02822}}.
	%%CITATION = ARXIV:1706.02822;%%
	
	\bibitem{Cornagliotto:2017dup}
	\nbbstauthor{M.~Cornagliotto, M.~Lemos and V.~Schomerus},
	\nbbsttitle{{Long Multiplet Bootstrap}},
	\nbbstjournal{\doiref{10.1007/JHEP10(2017)119}{JHEP~1710,~119~(2017)}},
	\nbbsteprint{\arxivref{1702.05101}{arxiv:1702.05101}}.
	%%CITATION = ARXIV:1702.05101;%%
	
	\bibitem{Kos:2018glc}
	\nbbstauthor{F.~Kos and J.~Oh},
	\nbbsttitle{{2d small N=4 Long-multiplet superconformal block}},
	\nbbstjournal{\doiref{10.1007/JHEP02(2019)001}{JHEP~1902,~001~(2019)}},
	\nbbsteprint{\arxivref{1810.10029}{arxiv:1810.10029}}.
	%%CITATION = ARXIV:1810.10029;%%
	
	\bibitem{Ramirez:2018lpd}
	\nbbstauthor{I.~A.~Ramírez},
	\nbbsttitle{{Towards general super Casimir equations for $4D$ ${\mathcal N}=1$
			SCFTs}},
	\nbbstjournal{\doiref{10.1007/JHEP03(2019)047}{JHEP~1903,~047~(2019)}},
	\nbbsteprint{\arxivref{1808.05455}{arxiv:1808.05455}}.
	%%CITATION = ARXIV:1808.05455;%%
	
	\bibitem{Dobrev:1977qv}
	\nbbstauthor{V.~K.~Dobrev, G.~Mack, V.~B.~Petkova, S.~G.~Petrova and
		I.~T.~Todorov},
	\nbbsttitle{{Harmonic Analysis on the n-Dimensional Lorentz Group and Its
			Application to Conformal Quantum Field Theory}},
	\nbbstjournal{\doiref{10.1007/BFb0009678}{Lect.~Notes~Phys.~63,~1~(1977)}}.
	%%CITATION = LNPHA,63,1;%%
	
	\bibitem{Quella:2007hr}
	\nbbstauthor{T.~Quella and V.~Schomerus},
	\nbbsttitle{{Free fermion resolution of supergroup WZNW models}},
	\nbbstjournal{\doiref{10.1088/1126-6708/2007/09/085}{JHEP~0709,~085~(2007)}},
	\nbbsteprint{\arxivref{0706.0744}{arxiv:0706.0744}}.
	%%CITATION = ARXIV:0706.0744;%%
	
	\bibitem{Kostant:1975qe}
	\nbbstauthor{B.~Kostant},
	\nbbsttitle{{Graded Manifolds, Graded Lie Theory, and Prequantization}},
	\nbbstjournal{\doiref{10.1007/BFb0087788}{Lect.~Notes~Math.~570,~177~(1977)}},
	in: \nbbsttitle{{Conference on Differential Geometrical Methods in Mathematical
			Physics Bonn, Germany, July 1-4, 1975}},
	pp.~177-306.
	%%CITATION = LNMAA,570,177;%%
	
	\bibitem{Blattner}
	\nbbstauthor{R.~J.~Blattner},
	\nbbsttitle{{Induced and Produced Representations of Lie Algebras}},
	\nbbstjournal{\doiref{10.2307/1995292}{Transactions~of~the~American~Mathematical~Society~144,~457~(1969)}}.
	
	\bibitem{Kac:1977em}
	\nbbstauthor{V.~G.~Kac},
	\nbbsttitle{{Lie Superalgebras}},
	\nbbstjournal{\doiref{10.1016/0001-8708(77)90017-2}{Adv.~Math.~26,~8~(1977)}}.
	%%CITATION = ADMTA,26,8;%%
	
	\bibitem{Schomerus:2005bf}
	\nbbstauthor{V.~Schomerus and H.~Saleur},
	\nbbsttitle{{The GL(1$|$1) WZW model: From supergeometry to logarithmic CFT}},
	\nbbstjournal{\doiref{10.1016/j.nuclphysb.2005.11.013}{Nucl.~Phys.~B734,~221~(2006)}},
	\nbbsteprint{\arxivref{hep-th/0510032}{hep-th/0510032}}.
	%%CITATION = HEP-TH/0510032;%%
	
	\bibitem{Saleur:2006tf}
	\nbbstauthor{H.~Saleur and V.~Schomerus},
	\nbbsttitle{{On the SU(2$|$1) WZW model and its statistical mechanics
			applications}},
	\nbbstjournal{\doiref{10.1016/j.nuclphysb.2007.02.031}{Nucl.~Phys.~B775,~312~(2007)}},
	\nbbsteprint{\arxivref{hep-th/0611147}{hep-th/0611147}}.
	%%CITATION = HEP-TH/0611147;%%
	
	\bibitem{Gotz:2006qp}
	\nbbstauthor{G.~Gotz, T.~Quella and V.~Schomerus},
	\nbbsttitle{{The WZNW model on PSU(1,1$|$2)}},
	\nbbstjournal{\doiref{10.1088/1126-6708/2007/03/003}{JHEP~0703,~003~(2007)}},
	\nbbsteprint{\arxivref{hep-th/0610070}{hep-th/0610070}}.
	%%CITATION = HEP-TH/0610070;%%
	
	\bibitem{messiah1962quantum}
	\nbbstauthor{A.~Messiah},
	\nbbsttitle{Quantum Mechanics},
	North-Holland Publishing Company (1962).
	
	\bibitem{Scheunert:1976wj}
	\nbbstauthor{M.~Scheunert, W.~Nahm and V.~Rittenberg},
	\nbbsttitle{{Irreducible Representations of the OSP(2,1) and SPL(2,1) Graded
			Lie Algebras}},
	\nbbstjournal{\doiref{10.1063/1.523149}{J.~Math.~Phys.~18,~155~(1977)}}.
	%%CITATION = JMAPA,18,155;%%
	
	\bibitem{Cordova:2016emh}
	\nbbstauthor{C.~Cordova, T.~T.~Dumitrescu and K.~Intriligator},
	\nbbsttitle{{Multiplets of Superconformal Symmetry in Diverse Dimensions}},
	\nbbsteprint{\arxivref{1612.00809}{arxiv:1612.00809}}.
	%%CITATION = ARXIV:1612.00809;%%
	
	\bibitem{Isachenkov:2017qgn}
	\nbbstauthor{M.~Isachenkov and V.~Schomerus},
	\nbbsttitle{{Integrability of conformal blocks. Part I. Calogero-Sutherland
			scattering theory}},
	\nbbstjournal{\doiref{10.1007/JHEP07(2018)180}{JHEP~1807,~180~(2018)}},
	\nbbsteprint{\arxivref{1711.06609}{arxiv:1711.06609}}.
	%%CITATION = ARXIV:1711.06609;%%
	
	\bibitem{Vilenkin}
	\nbbstauthor{N.~Ja.~Vilenkin and A.~U.~Klimyk},
	\nbbsttitle{Representation of Lie Groups and Special Functions},
	Springer Netherlands (1991).
	
	\bibitem{Reshetikhin:2015rba}
	\nbbstauthor{N.~Reshetikhin},
	\nbbsttitle{{Degenerate integrability of quantum spin Calogero–Moser
			systems}},
	\nbbstjournal{\doiref{10.1007/s11005-016-0897-8}{Lett.~Math.~Phys.~107,~187~(2017)}},
	\nbbsteprint{\arxivref{1510.00492}{arxiv:1510.00492}}.
	%%CITATION = ARXIV:1510.00492;%%
	
	\bibitem{Reshetikhin}
	\nbbstauthor{N.~Reshetikhin},
	\nbbsttitle{{Spin Calogero-Moser models on symmetric spaces}},
	\nbbsteprint{\arxivref{1903.03685}{arxiv:1903.03685}}.
	
	\bibitem{Kostant}
	\nbbstauthor{B.~Kostant and J.~Tirao},
	\nbbsttitle{On the Structure of Certain Subalgebras of a Universal Enveloping
		Algebra},
	\nbbstjournal{\doiref{10.1090/S0002-9947-1976-0404367-0}{Transactions~of~The~American~Mathematical~Society~218,~133~(1976)}}.
	
	\bibitem{Billo:2016cpy}
	\nbbstauthor{M.~Billò, V.~Gonçalves, E.~Lauria and M.~Meineri},
	\nbbsttitle{{Defects in conformal field theory}},
	\nbbstjournal{\doiref{10.1007/JHEP04(2016)091}{JHEP~1604,~091~(2016)}},
	\nbbsteprint{\arxivref{1601.02883}{arxiv:1601.02883}}.
	%%CITATION = ARXIV:1601.02883;%%
	
	\bibitem{Lauria:2017wav}
	\nbbstauthor{E.~Lauria, M.~Meineri and E.~Trevisani},
	\nbbsttitle{{Radial coordinates for defect CFTs}},
	\nbbstjournal{\doiref{10.1007/JHEP11(2018)148}{JHEP~1811,~148~(2018)}},
	\nbbsteprint{\arxivref{1712.07668}{arxiv:1712.07668}}.
	%%CITATION = ARXIV:1712.07668;%%
	
	\bibitem{Lauria:2018klo}
	\nbbstauthor{E.~Lauria, M.~Meineri and E.~Trevisani},
	\nbbsttitle{{Spinning operators and defects in conformal field theory}},
	\nbbsteprint{\arxivref{1807.02522}{arxiv:1807.02522}}.
	%%CITATION = ARXIV:1807.02522;%%
	
	\bibitem{Balitsky:2013npa}
	\nbbstauthor{I.~Balitsky, V.~Kazakov and E.~Sobko},
	\nbbsttitle{{Two-point correlator of twist-2 light-ray operators in N=4 SYM in
			BFKL approximation}},
	\nbbsteprint{\arxivref{1310.3752}{arxiv:1310.3752}}.
	%%CITATION = ARXIV:1310.3752;%%
	
	%\cite{Balitsky:2015tca}
         \bibitem{Balitsky:2015tca}
         I.~Balitsky, V.~Kazakov and E.~Sobko,
         ``Structure constant of twist-2 light-ray operators in the Regge limit,''
          Phys.\ Rev.\ D {\bf 93} (2016) no.6,  061701
          doi:10.1103/PhysRevD.93.061701
          \nbbsteprint{\arxivref{1506.02038}{arxiv:1506.02038}}.
          
          
          %\cite{Balitsky:2015oux}
           \bibitem{Balitsky:2015oux}
           I.~Balitsky, V.~Kazakov and E.~Sobko,
           ``Three-point correlator of twist-2 light-ray operators in N=4 SYM in BFKL approximation,''
            \nbbsteprint{\arxivref{1511.03625}{arxiv:1511.03625}}.
           
	
\end{thebibliography}
\end{document}